\documentclass[useAMS]{mn2e}
\usepackage{amsmath}
\usepackage{textcomp}
\usepackage{amssymb}
\usepackage[pdftex]{graphicx}
\usepackage{amssymb}

\usepackage[margin=.8in, paperwidth=8.5in, paperheight=11in]{geometry}
\usepackage{multirow}
\usepackage{multicol}
\usepackage{epsfig}
\usepackage{graphicx}
\usepackage{makecell}
\usepackage{array}
\usepackage{tikz}
\usetikzlibrary{decorations.pathreplacing,calc}

\newcommand{\tikzmark}[2][-3pt]{\tikz[remember picture, overlay, baseline=-0.5ex]\node[#1](#2){};}

\tikzset{brace/.style={decorate, decoration={brace}},
 brace mirrored/.style={decorate, decoration={brace,mirror}},
}

\newcounter{brace}
\newcommand{\drawbrace}[3][brace]{%
 \refstepcounter{brace}
 \tikz[remember picture, overlay]\draw[#1] (#2.center)--(#3.center)node[pos=0.5, name=brace-\thebrace]{};
}

\newcounter{arrow}
\newcommand{\annote}[3][]{%
 \tikz[remember picture, overlay]\node[#1] at (#2) {#3};
}

\newcommand\ddfrac[2]{\frac{\displaystyle #1}{\displaystyle #2}}

\begin{document}

\title{Lense-Thirring Precession of Misaligned Discs I}
\author[S. Dyda,C.S. Reynolds]
{\parbox{\textwidth}{Sergei~Dyda\thanks{sdyda@ast.cam.ac.uk}, Christopher S. Reynolds}\\
Institute of Astronomy, Madingley Road, Cambridge CB3 0HA, UK
}

\date{\today}
\pagerange{\pageref{firstpage}--\pageref{lastpage}}
\pubyear{2020}

\label{firstpage}

\maketitle

\begin{abstract}
We study Lense-Thirring precession of inviscid and viscous misaligned $\alpha-$discs around a black hole using a gravitomagnetic term in the momentum equation. For weak misalignments, $i \lesssim 10^{\circ}$, the discs behave like rigid bodies, undergoing the full suite of classical harmonic oscillator dynamics including, weak and critically damped motion (due to viscosity), precession (due to Lense-Thirring torque) and nutation (due to apsidal precession). For strong misalignments, $i \gtrsim 30^{\circ}$, we find sufficiently thin, $h/r \lesssim 0.05$ discs break, form a gap and the inner and outer sub-discs evolve quasi independently apart from slow mass transfer. Assuming the sound speed sets the communication speed of warps in the disc, we can estimate the breaking radius by requiring that the inner sub-disc precesses like a rigid body.  We explicitly show for the first time using a grid code that an Einstein potential is needed to reproduce the analytic properties of the inner disc edge and find disc breaking. At large inclination angles we find multiple disc breaks, consistent with recent GRMHD simulations of highly inclined discs. Our results suggest that the inclusion of a gravitomagnetic term and appropriate pseudo-Newtonian potential captures the important quantitative features of misaligned discs.

\end{abstract}


\section{Introduction}

Accretion discs can form at late times in a black holes formation history, via accretion events or galaxy mergers (Volonteri et al. 2005; King et al. 2005), producing a system where the orientation of the spin angular momentum of the black hole and the orbital angular momentum of the disc are \emph{misaligned}. In such a scenario, the dynamics are expected to be strongly influenced by general relativitic (GR) effects.  Local frame-dragging, associated with black hole spin, induces Lense-Thirring precession (Lense \& Thirring 1918) and the precession coupled to viscous effects lead to Bardeen-Petterson alignment of the inner disc (Bardeen \& Petterson 1975, hereafter BP75). Understanding how the disc dynamics and resulting observables are affected by GR can provide information on the accretion flow near the ISCO which may improve black hole spin measurements (see Reynolds 2019 for a review) and possibly explain phenomena such as low frequency quasi-periodic oscillations (QPOs) (Ingram et al. 2009). 

Analytic studies of disc warps considered two different regimes, based on the relative size of the dimensionless viscosity parameter $\alpha$ and disc half-thickness $h/r$. In the \emph{diffusive} regime, $\alpha \gtrsim h/r$, the evolution of warps are described by a diffusion equation (Papaloizou \& Pringle 1983) and angular momentum transport mediated by the disc viscosity leads to alignment of the inner disc up to the Bardeen-Petterson radius $r_{\rm{BP}}$ (Kumar \& Pringle 1985; Pringle 1992). Further work by Ogilvie (1999, 2000 hereafter O99/00 respectively) extended this theory to the non-linear regime, to describe discs with warps of arbitrary size. Alternatively, in the \emph{bending-wave} regime, $\alpha \lesssim h/r$, warps propagate as waves at half the speed of sound (Papaloizou \& Lin 1995) and radial tilt oscillations occur near the inner edge of the disc in a Kerr geometry (Ivanov \& Illarionov 1997; Lubow, Ogilvie \& Pringle 2002). Unlike in the diffusive regime, a non-linear theory of warped discs in the bending-wave regime has so far remained elusive.

Despite the progress made in studying warped discs analytically, the challenging nature of this problem lends itself well to investigation using numerical simulations. Nelson and Papaloizou (2000) used smooth particle hydrodynamics (SPH) simulations in both the diffusive and wave regimes to study warped discs driven by a Lense-Thirring torque and found evidence of Bardeen-Petterson alignment. Later simulations by Lubow, Ogilvie and Pringle (2002) did not agree with these simulations, as they found radial tilt oscillations disrupted the inner disc alignment. SPH simulations by Lodato and Price (2010) showed that disc warps in isolated discs are well described by the non-linear analytic theory O99/00.  However, in the case of discs driven by a Lense-Thirring torque, Nealon et al. (2015) showed that unlike in the analytic theory where warps remain smooth, discs can break into discrete rings that then precess independently.    

Grid based MHD codes have also been used to study this problem, beginning with Sorathia et al. (2013a) which studied the relaxation of an unforced warp in inviscid hydrodynamics. In their follow up work, (Sorathia et al. 2013b), they compared the evolution of an inviscid and viscous discs under an external Lense-Thirring torque and found that though magnetic forces are small compared to internal pressure forces, they can alter the propagation of waves in the disc and play a key role in the inner disc aligning. Follow up work by Krolik \& Hawley (2015) attempted to understand where this transition occurs by loading additional matter in the outer parts of the disc and studying how the alignment front propagates outward.  Krolik \& Hawley (2018) showed that the alignment properties are only weakly dependent on sound speed. Krolik \& Hawley (2019) studied disc alignment for moderately inclined discs, $i \leq 24^{\circ}$, in a Newtonian potential and found discs align at a nearly uniform rate but do not break. 

Rather than incorporating the effects of a Kerr black hole with misaligned spin via an external Lense-Thirring torque term and pseudo-Newtonian potential, some groups have made use of full GRMHD codes where these effects are included by directly solving the GR equations of motion. Fragile \& Anninos (2005) performed the first such simulations using the \textsc{COSMO} code and found misaligned thick discs form a warp and precess nearly like a rigid body. Follow up work by Fragile et al. (2007) showed that these results held when the effects of the MRI were included, though this effect was marginally resolved. These early simulations, along with later better resolved models (Morales Teixeira et al. 2014; Zhuravlev et al. 2014) found no evidence of Bardeen-Petterson alignment, though the latest simulations of very thin discs, $h/r = 0.03$, do find alignment very close to the black hole (Liska et al. 2019). Further, at high inclination angles, the disc is found to tear (Liska et al. in prep). White, Quataert \& Blaes (2019) conducted a systematic survey of inclined discs for a range of spins $0 \leq a \leq 0.9$ at low inclination $i \leq 24^{\circ}$.  The already complex dynamics of a misaligned disc can be further complicated by the addition of a relativistic jet which can further align the inner disc edge by providing a magnetic torque (Polko \& McKinney 2017) or by altering the rate of precession (Liska et al. 2018). 

Each of the aforementioned numerical methods, SPH, grid based MHD and GRMHD offers its own strengths and potential drawbacks. Strong evidence suggests that astrophysical discs are effectively viscous due to the generation of Maxwell stresses via the MRI (Balbus \& Hawley 1991). The simplest possible treatment of this viscosity is via a Shakura-Sunyaev (1973), $\alpha-$disc parametrization. SPH simulations are computationally inexpensive and can explore a range of viscosity parameters in $\alpha-$disc models. Though this is a good first step, it suffers from two drawbacks. Firstly, both shearing box and global disc simulations have shown that the viscous stress is not uniform and isotropic as in the $\alpha-$disc picture. Secondly, numerical viscosity in SPH simulations is typically \emph{larger} than in physical accretion discs, being set by the particle number so computational restrictions set a lower bound on disc viscosity in low density regions. SPH simulations are thus unable to study inviscid discs as well as discs where viscosity is generated self consistently via the MRI. This may be important for this problem since disc viscosity plays a key role in the disc evolution and disc simulations where the MRI is treated self-consistently have shown that the effective viscosity parameter $\alpha$ is highly non-uniform. Finally the GR effects, which drive the dynamics of interest, are only implemented at lowest order, by including an additional source term in the momentum equation and using a pseudo-Newtonian gravitational potential. This limitation may be overcome in the future however using new GRSPH codes (Liptai \& Price 2019) which capture the kinematics of particles in curved space-time. 

Grid based MHD simulations are computationally more expensive than SPH, particularly for large misalignment angles which require enhanced resolution for a wide range of azimuthal angles. Isotropic viscosity can be implemented for a range of $\alpha$ values down to the grid viscosity scale. With sufficiently high resolution, $\sim 16$ grid cells per scale height the MRI can be resolved and viscosity calculated self consistently. In essentially Newtonian treatments the GR effects are implemented as in SPH with an Lense-Thirring source term and pseudo-Newtonian potential.   

GRMHD is computationally the most expensive of all methods, despite recent advances of optimizing codes for using GPUs (Liska et al. 2018). They can use an isotropic viscosity, though in practice groups have either studied inviscid discs or those where the viscosity is generated via the MRI. There is no need to implement any additional source terms due to spin-orbit coupling or an effective gravitational potential as these codes solve the GR equations of motion from which these effects are derived. GRMHD thus provides the most physically accurate modeling of misaligned systems but at high computational cost. Further because GR effects are all fully included it can be difficult to interpret results as different physical effects cannot be disentangled.

In this series of papers we study the Lense-Thirring precession of inclined accretion discs. We use the grid based code \textsc{Athena++} (Stone et al. 2020) to simulate an accretion disc misaligned with respect to the black hole spin. We use the Newtonian HD module, and include the effects of GR by imposing an external torque derived from the lowest order gravitomagnetic correction of spin-orbit coupling and an effective GR gravitational potential that correctly reproduces the apsidal precession frequency at large radii. We take the approach that grid based HD simulations allow us to methodologically add additional physical effects (Lense-Thirring torque, GR effective potential, $\alpha$ viscosity; Paper I) and MRI generated viscosity (Paper II). The low computational cost, relative to GRMHD, allows us to more fully explore the relevant parameter space (misalignment angle $i$, viscosity parameter $\alpha$, disc half thickness $h/r$) while maintaining sufficient grid resolution. Our goal is to qualitatively understand the different evolutionary scenarios (disc precession, nutation and breaking) and quantify physically relevant quantities (apsidal and nodal precession frequencies and breaking radius) which we may be able to correlate with observables such as iron line reflection spectra and QPOs.

\section{Theory}     
We study precession of inclined accretion discs by including first order GR corrections to the classical HD equations. This includes an external torque induced from the spin-orbit coupling between the black hole and fluid, the so called Lense-Thirring term (nodal precession), and a modified gravitational potential (apsidal precession). To further simplify the problem we assume the disc is isothermal. Below we describe the basic equations (Section \ref{sec:equations}) and initial and boundary condition (Section \ref{sec:InitCond}). Interested readers can find details of our initial setup, including coordinate transformations into the inclined discs frame (Appendix \ref{sec:app_ic}) and implementation of the Lense-Thirring term into \textsc{Athena++} (Appendix \ref{sec:LenseThirring}). 
    
\subsection{Basic Equations}
\label{sec:equations}

The basic equations for isothermal single fluid hydrodynamics with GR corrections are
\begin{subequations}
\begin{equation}
\frac{\partial \rho}{\partial t} + \nabla \cdot \left( \rho \mathbf{v} \right) = 0,
\end{equation}
\begin{equation}
\frac{\partial (\rho \mathbf{v})}{\partial t} + \nabla \cdot \left(\rho \mathbf{vv} + \mathbf{P} + \mathsf{\tau} \right) = - \rho \nabla \Phi + \rho \left( \mathbf{v} \times \mathbf{h} \right),
\label{eq:momentum}
\end{equation}
\label{eq:hydro}%
\end{subequations}
where $\rho$ is the fluid density, $\mathbf{v}$ the velocity, $\mathbf{P}$ a diagonal tensor with components $P = \rho c_s^2$ the gas pressure, $c_s$ is the isothermal sound speed and $\mathsf{\tau}$ is the viscosity tensor. The gravitomagnetic vector field is
\begin{equation}
\mathbf{h} = \frac{2\mathbf{J}}{r^3} - \frac{6 \left( \mathbf{J} \cdot \mathbf{r} \right) \mathbf{r}}{r^5},
\label{eq:gravitomagnetic}
\end{equation}
where $r$ is the spherical radius and the spin vector of the black hole $\mathbf{J} = a (GM)^2/c^3 \ \hat{z}$ where $M$ and $a$ are the black hole mass and dimensionless spin parameter respectively. Further details of our numerical implementation of this source term can be found in Appendix \ref{sec:LenseThirring}.

We consider two possible forms of the gravitational potential of the central object $\Phi$: the usual Newtonian potential
\begin{equation}
\Phi_{\rm{N}} = - \frac{GM}{r},
\label{eq:NewtonPotential}
\end{equation}
and an effective GR potential, (Nelson \& Papaloizou 2000)
\begin{equation}
\Phi_{\rm{GR}} = - \frac{GM}{r}\left[ 1 + \frac{3 r_g}{r}\right],
\label{eq:GRPotential}
\end{equation}
with $r_g = GM/c^2$ the gravitational radius. The potential (\ref{eq:GRPotential}) correctly reproduces the correct GR apsidal precession rate at large radii.

We model the viscosity using the Shakura-Sunyaev $\alpha-$disc prescription, where the kinematic viscosity is given by
\begin{equation}
\nu = \alpha_{\nu} \frac{c_s^2}{\Omega_K},
\end{equation}
for dimensionless parameter $0 \leq \alpha_{\nu} \leq 10^{-2}$ and $\Omega_K$ the Keplerian orbital frequency.

\subsection{Numerical Setup}
\label{sec:InitCond}

The central object has a mass $M = 10 M_{\odot}$ and has gravitational radius $r_g = 1.47 \times 10^{6} \ \rm{cm}$. The simulation region extends from $[r_{\rm{in}},r_{\rm{out}}] = \left\{4 r_g \leq r \leq 40 r_g\right\}$. We express our results using units of length in $r_g$ and time in inner disc orbital periods $T_0 = 2 \pi (r_{\rm{in}}/r_g)^{3/2} GM/c^2$.  We use a logarithmically spaced grid of $N_r = 128$ points and a scale factor $a_r = 1.02$ that defines the grid spacing recursively via $dr_{n+1} = a_r dr_n$. We choose a uniform grid in the azimuthal and axial directions with $N_{\theta} = 192$ and $N_{\phi} = 192$ points respectively, spanning $0.1 \leq \theta \leq \pi - 0.1$ and $0 \leq \phi \leq 2 \pi$. 

Our initial setup is a disc in vertical hydrostatic balance, inclined by an angle $10^{\circ} \leq i \leq 45^{\circ}$ relative to the black hole with dimensionless spin $a = 0.9$. The disc has Keplerian velocity on cylinders oriented with the disc at fixed radial distance from the black hole and has initially constant $0.025 \leq h/r \leq 0.1 $. To thermally launch outflows, the hydrodynamic escape parameter $\rm{HEP} = GM/r_{\rm{in}} c_s^2 \lesssim 10$ so for our choice of disc half-thicknesses it falls in the range $ 25 \leq \rm{HEP} \leq 400 $ so in this sense the disc can be thought of as cold. 

At the inner and outer radial boundaries we impose outflow boundary conditions. We use axis boundary conditions along the axial boundaries and periodic conditions in $\phi$. Further details of our setup can be found in Appendix \ref{sec:app_ic}.

\subsection{Precession Frequencies}
\label{sec:frequencies}
Defining the right hand side of (\ref{eq:momentum}) as an effective potential and using the Newtonian definitions of epicyclic and vertical frequencies one can compute the apsidal and nodal precession frequencies (see Nealon et al. 2015, equations (27)-(30)). For the Newtonian potential (\ref{eq:NewtonPotential}) these are respectively, in our dimensionless code units 
\begin{subequations}
\begin{equation}
\eta_N = 3 \pi r_{\rm{in}}^{3/2} \frac{a}{r^3}\frac{1}{1 - 2 a r^{-3/2}},
\label{eq:N_eta}
\end{equation}
\begin{equation}
\xi_N = 4 \pi r_{\rm{in}}^{3/2} \frac{a}{r^3}\frac{1}{1 - 2 a r^{-3/2}}.
\label{eq:N_xi}
\end{equation} 
\label{eq:N_f}
\end{subequations}
Likewise for the GR potential (\ref{eq:GRPotential})
\begin{subequations} 
\begin{equation}
\eta_{\rm{GR}} = \frac{\Omega_{\rm{GR}}}{2} \frac{\left[ 6 r^{-1} - 3ar^{-3/2}\right]}{\left[ 1 + 6 r^{-1} - 2ar^{-3/2} \right]}, 
\label{eq:GR_eta}
\end{equation}
\begin{equation}
\xi_{\rm{GR}} = - 2a  \frac{\Omega_{\rm{GR}}}{r^{3/2} \left[ 1 + 6 r^{-1} - 2ar^{-3/2} \right]},
\label{eq:GR_xi}
\end{equation}
\label{eq:GR_f}
\end{subequations}
where the GR corrected orbital frequency
\begin{equation}
\Omega_{\rm{GR}} = 2 \pi \left(\frac{r_{\rm{in}}}{r}\right)^{3/2} \left[ 1 + \frac{6}{r} - \frac{2a}{r^{3/2}}\right]^{1/2}.
\end{equation}
In parts of our analysis we treat the disc like a rigid body and define angular momentum averaged precessional frequencies
\begin{equation}
\bar{f}_{\Phi} = \ddfrac{\int_{r_1}^{r_2} f_{\Phi} L_{\perp} \ dr}{\int_{r_1}^{r_2} L_{\perp} \ dr},
\label{eq:fbar}
\end{equation}
for $f = \eta$ or $\xi$ and $\Phi$ the appropriate potential.  

\section{Results}

\begin{figure*}
                \centering
                \includegraphics[width=\textwidth]{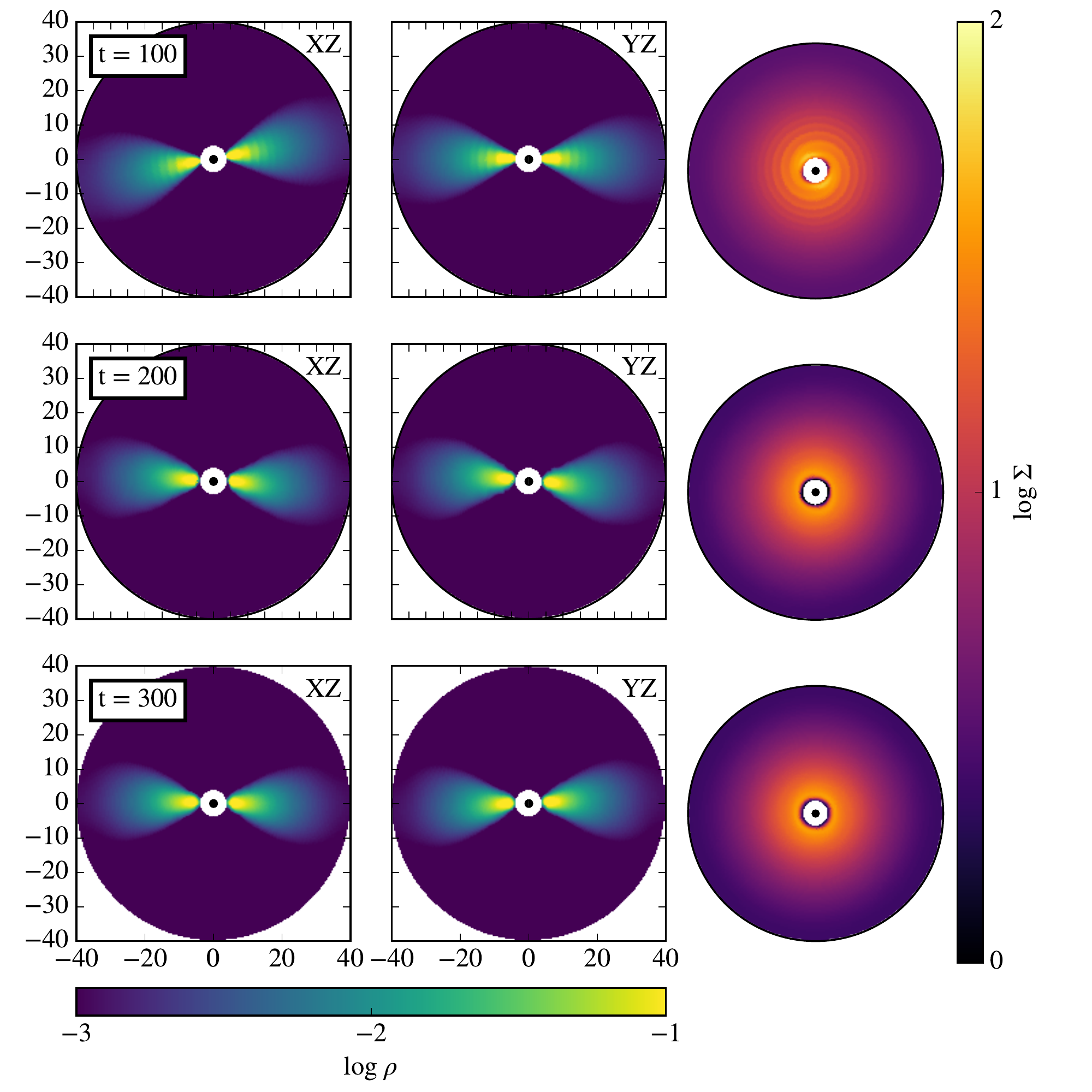}
        \caption{Density $\rho$ in the XZ and YZ planes and surface density $\Sigma$ for $i=10^{\circ}$ Newtonian disc at t = 0, 100 and 200. The disc undergoes nearly rigid body precession and aligns into the black hole spin midplane.}
\label{fig:summary_i10}
\end{figure*} 

\begin{table*}
\begin{center}
\begin{tabular}{ l |l| c  c  c| c| c c c c | l}
\multicolumn{1}{c}{$i$} 	& \multicolumn{1}{l}{Model} 	& \multicolumn{1}{c}{$\alpha$} 	& \multicolumn{1}{c}{$\Phi$} & \multicolumn{1}{c}{$h/r$}	& \multicolumn{1}{c}{$R$}	&\multicolumn{1}{c}{$\bar{\eta}$}& \multicolumn{1}{c}{$\bar{\xi}$}	& \multicolumn{1}{c}{$\omega_{\eta}$} & \multicolumn{1}{c}{$\omega_{\xi}$} & \\ \hline \hline
\tikzmark[xshift=-8pt,yshift=1ex]{x} \multirow{4}{*}{$10$} & Newtonian	& 0		& N 	& 0.1 	&$\lesssim 40.0$ & 0.04	& 0.05	& 0.09	& 0.04	& Underdamped rigid body precession \\
 	& Inviscid   & 0		& GR 	& 0.1 	&$\lesssim 40.0$ &0.22	& 0.05	& 0.09	&0.03, 0.06& Underdamped precession \& nutation \\
	& Low Visc.  & $10^{-3}$	& GR 	& 0.1 	&$\lesssim 40.0$ &0.13	& 0.03	& 0.12	&0.03, 0.09& Underdamped precession \& nutation \\ 
\tikzmark[xshift=-8pt,yshift=-1ex]{y} 	& High Visc. & $10^{-2}$	& GR 	& 0.1 	&$\lesssim 40.0$ &0.13	& 0.03	& 0.16	& 0.13	& Critically damped precession \\ \hline
\tikzmark[xshift=-8pt,yshift=1ex]{z}\multirow{10}{*}{$30$}	& \multirow{2}{*}{Newtonian} & \multirow{2}{*}{0}		& \multirow{2}{*}{N} 	& \multirow{2}{*}{0.05} 	& $\lesssim 12.5$  		 & 0.46	& 0.12	& 0.12	& 0.12	& \multirow{2}{*}{\makecell[l]{Disc break. Inner BP \\ alignment and outer precession}}  \\
	& & & &  & $\gtrsim 12.5$		 & 0.47	& 0.12	& 0.16	& 0.16	&   \\ 
\multirow{2}{*}{} 	& \multirow{2}{*}{Inviscid} & \multirow{2}{*}{0}		& \multirow{2}{*}{GR} 	& \multirow{2}{*}{0.05} 	& $\lesssim 9.0$ & 0.75	& 0.21	& -	& 0.28	& \multirow{2}{*}{Disc break. Inner and outer precession} \\  
&  & & &   		& $\gtrsim 11.1$ & 0.06	& 0.009	& 0.14	& 0.015	& \\ 
 	& \multirow{2}{*}{Low Visc.}& \multirow{2}{*}{$10^{-3}$}	& \multirow{2}{*}{GR} 	& \multirow{2}{*}{0.05} 	& $\lesssim 9.9$& 0.90	& 0.27	& 0.39	& 0.39	& \multirow{2}{*}{Disc break. Inner and outer precession}\\ 
& & & &                    & $\gtrsim 11.1$ & 0.06	& 0.010	& -	& 0.015	& \\ 
	& \multirow{2}{*}{High Visc.} & \multirow{2}{*}{$10^{-2}$}	& \multirow{2}{*}{GR} 	& \multirow{2}{*}{0.05} & $\lesssim 6.6$ & 1.31	& 0.43	& -	& 0.63	& \multirow{2}{*}{\makecell[l]{Short lived inner disc which accretes and  \\ disc re-forms at late times and precesses}} \\
&  & & & & $\lesssim 23.0$	& 0.10	& 0.017	& -	& 0.020	& \\ 
 	& \multirow{2}{*}{Thin} & \multirow{2}{*}{0}		& \multirow{2}{*}{GR} 	& \multirow{2}{*}{0.025} & $\lesssim 12.5$ & 0.46& 0.12	& 0.12	& 0.12	& \multirow{2}{*}{Disc break. Inner and outer precession} \\ 
& & & &                    & $\gtrsim 12.5$ & 0.04	& 0.005	& -	& 0.008	& \\
 \multirow{2}{*}{$45$} & \multirow{2}{*}{High Inc.}	& \multirow{2}{*}{0}		& \multirow{2}{*}{GR} 	& \multirow{2}{*}{0.05}  & $\lesssim 5.4$ & 1.73	& 0.61	& -	& 1.01	& \multirow{2}{*}{\makecell[l]{Two disc breaks, all components precess. \\ Final state resembles low inclination disc}} \\ 
\tikzmark[xshift=-8pt,yshift=-1ex]{w}& & & &                    & $\gtrsim 12.5$ & 0.47	& 0.12	& 0.16	& 0.14	& \\ \hline\hline
    \end{tabular}
\drawbrace[brace mirrored, thick]{x}{y}
\annote[left]{brace-1}{\rotatebox{90}{Weak}}
\drawbrace[brace mirrored, thick]{z}{w}
\annote[left]{brace-2}{\rotatebox{90}{Strong}}
\end{center}
\caption{Summary of all models, including weakly ($i = 10^{\circ}$) and strongly ($i \gtrsim 30^{\circ}$) misaligned discs. We list the viscosity $\alpha$, gravitational potential $\Phi$ and disc half-thickness $h/r$. For each model we list the radial range R over which the listed apsidal $\eta$ and nodal $\xi$ frequencies are measured. We list both the theoretical frequencies $\bar{\eta}$ and $\bar{\xi}$ expected from equations (\ref{eq:N_f}) and (\ref{eq:GR_f}) respectively and from the Fourier analysis of the disc angular momentum.}
\label{tab:summary}
\end{table*}

We perform a series of numerical simulations to study the evolution of misaligned accretion discs. We consider two qualitatively different physical regimes: weakly misaligned discs (Section \ref{sec:low_inc}), which undergo nodal and apsidal precession, and strongly misaligned discs (Section \ref{sec:high_inc}) which in addition to the aforementioned physical effects break into nearly uncoupled inner and outer subdiscs. A summary of all our runs, as well as their most relevant parameters, is provided in Table \ref{tab:summary}. 

\subsection{Weakly Misaligned Discs}
\label{sec:low_inc}

We first consider the evolution of weakly misaligned, $i = 10^{\circ}$, accretion disc. The small inclination angle makes the disc evolution less extreme so we use this as a controlled test case. We explore the effects of the gravitational potential (Section \ref{sec:low_potentials}) and viscosity (Section \ref{sec:low_viscosity}).

The simplest case is an inviscid disc in a purely Newtonian potential (\ref{eq:NewtonPotential}) but subject to Lense-Thirring torques. We use this case to establish our main analysis tools and to benchmark further simulations. The disc is first allowed to reach a stationary state during an initial period of $t = 100$ inner disc orbits without any Lense-Thirring torques. The Lense-Thirring term is then turned on and after an initial transient spiral density wave beginning from the inner edge and propagating outward, the disc begins to undergo rigid body precession and then aligns with the midplane. In Fig. \ref{fig:summary_i10} we plot the density in the XZ and YZ planes as well as the disc surface density $\Sigma$ at representative times in the system evolution, $t = 100$, $200$ and $300$. 

To quantify the evolution we divide the disc into rings of fixed radius $r$ and define the angular momentum of each ring $\mathbf{l}(r) = (l_x,l_y,l_z)$. The total disc angular momentum is then
\begin{equation}
\mathbf{L} = \int_{r_{\rm{in}}}^{r_{\rm{out}}} \mathbf{l}(r) dr = \int_{r_{\rm{in}}}^{r_{\rm{out}}} \rho \:\mathbf{r} \times \mathbf{v} \ r^2 \sin \theta  d\theta d \phi dr.
\end{equation}
In addition, we define the discs angular momentum perpendicular to the black hole spin axis 
\begin{equation}
L_{\perp} = \sqrt{L_x^2 + L_y^2}.
\end{equation}
The angular momentum vectors allow us to define the alignment angle of each ring (see for example BP75) 
\begin{equation}
\beta = \tan^{-1} \left( \frac{l_{\perp}}{l_z}\right),
\end{equation}
and the precession angle
\begin{equation}
\gamma = \tan^{-1} \left( \frac{l_{y}}{l_x}\right), 
\end{equation}
where we choose the appropriate branches so a precessing disc will have $0 \leq \gamma < 2 \pi$.   

\begin{figure*}
                \centering
                \includegraphics[width=0.48\textwidth]{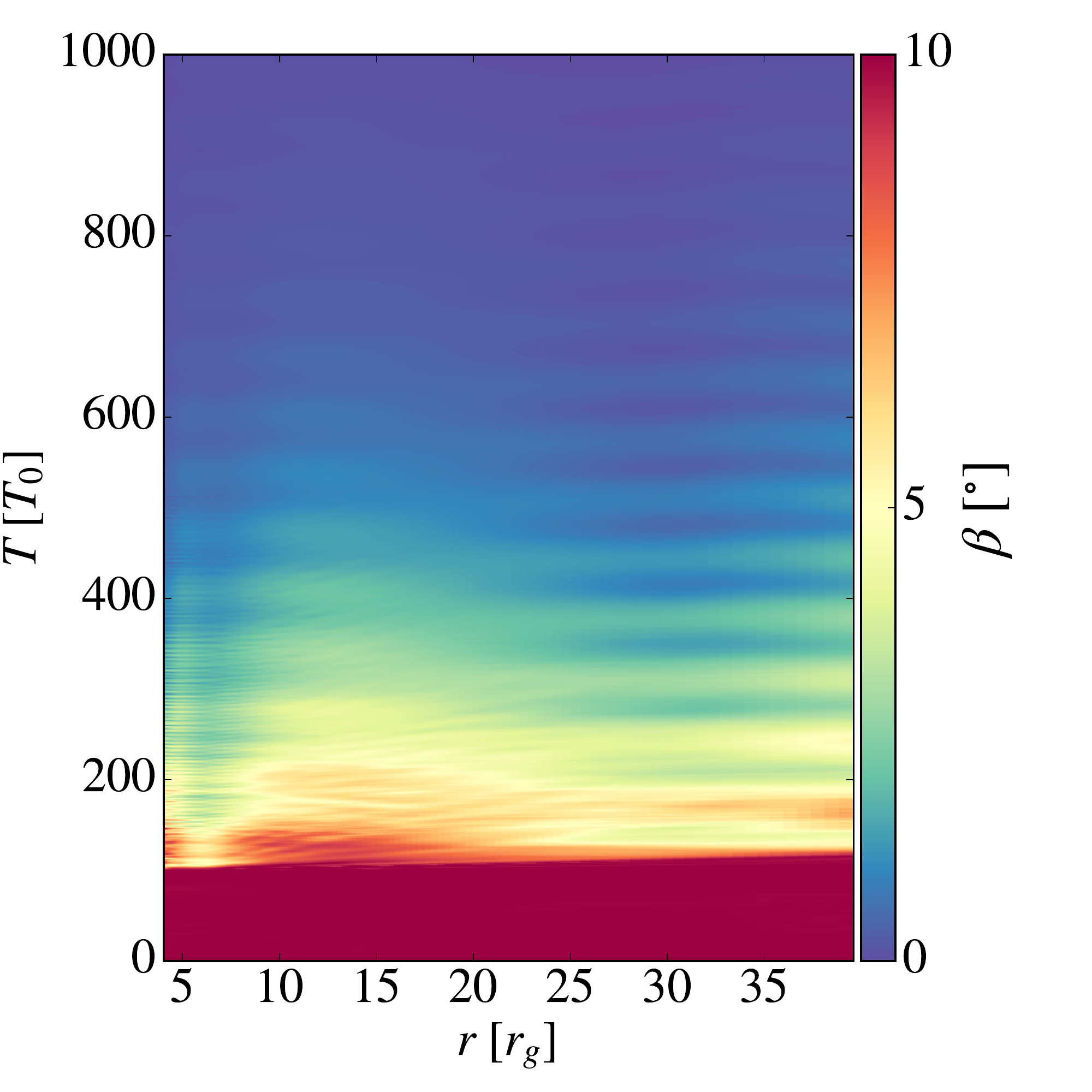}
                \includegraphics[width=0.48\textwidth]{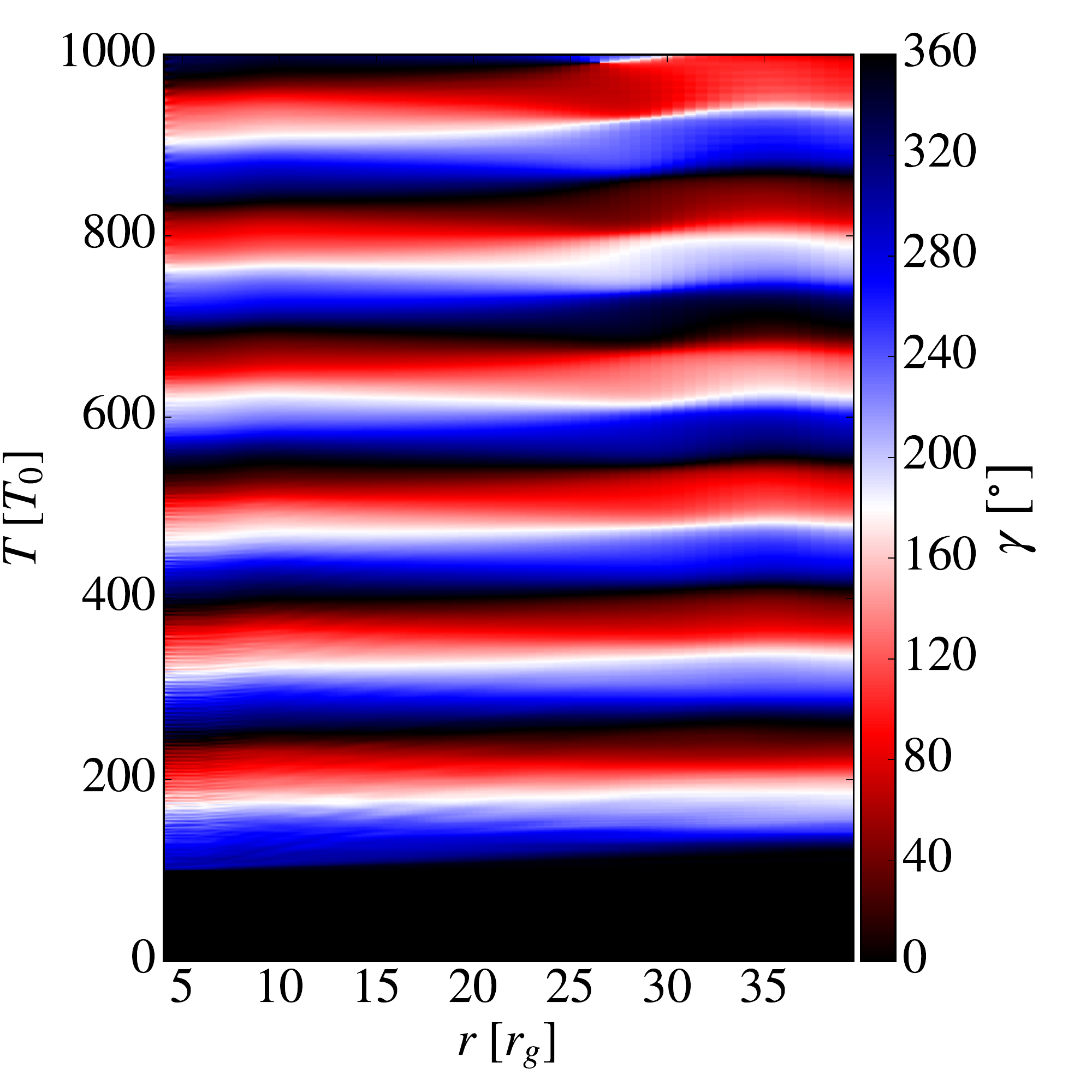}
        \caption{\textit{Left Panel -} Misalignment angle $\beta$ as a function of time and radius for the Newtonian disc \textit{Right Panel -} Precession angle $\gamma$ as a function of time and radius for the Newtonian disc.}
\label{fig:angles}
\end{figure*} 

\begin{figure}
                \centering
                \includegraphics[width=0.48\textwidth]{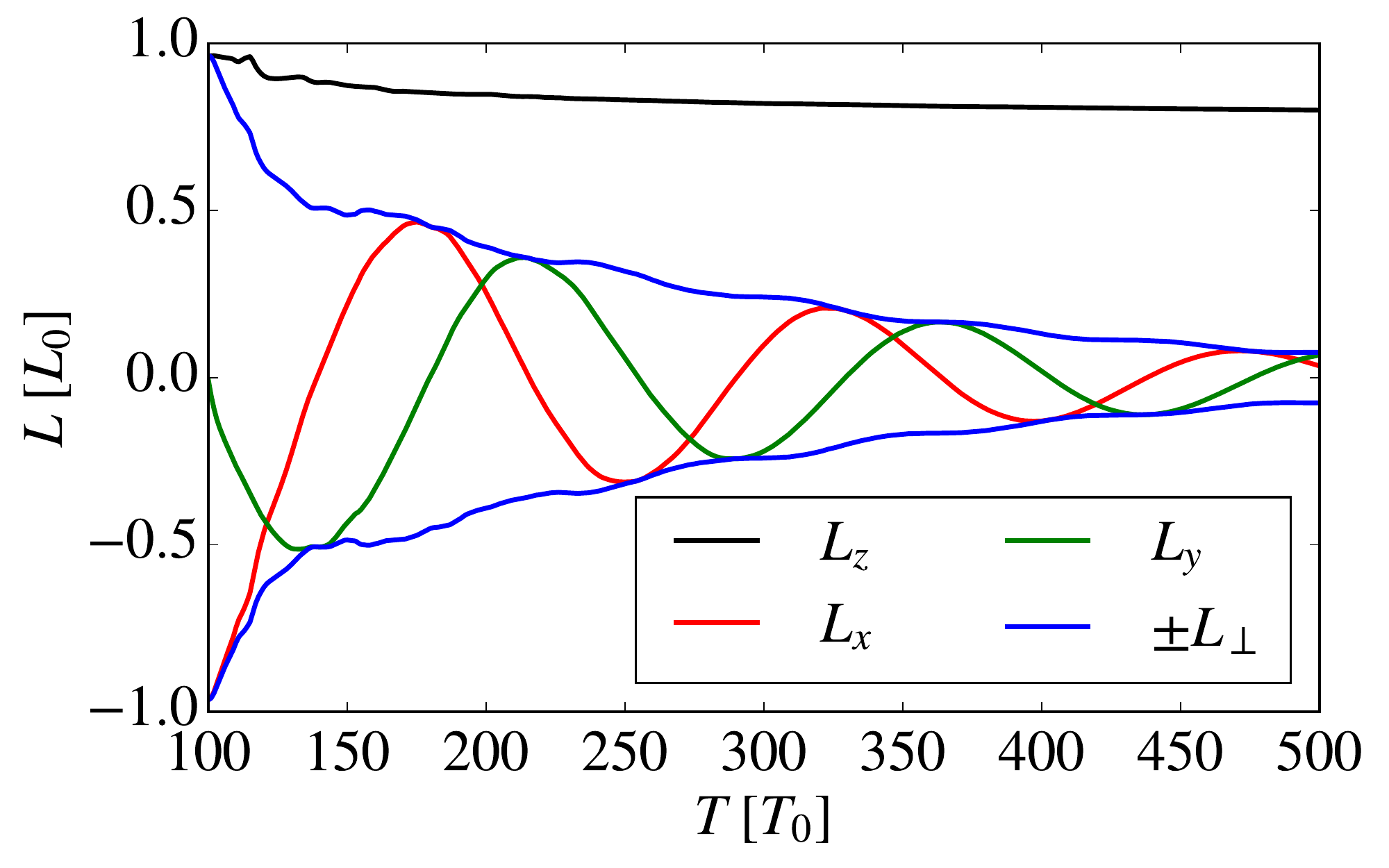}
                \includegraphics[width=0.48\textwidth]{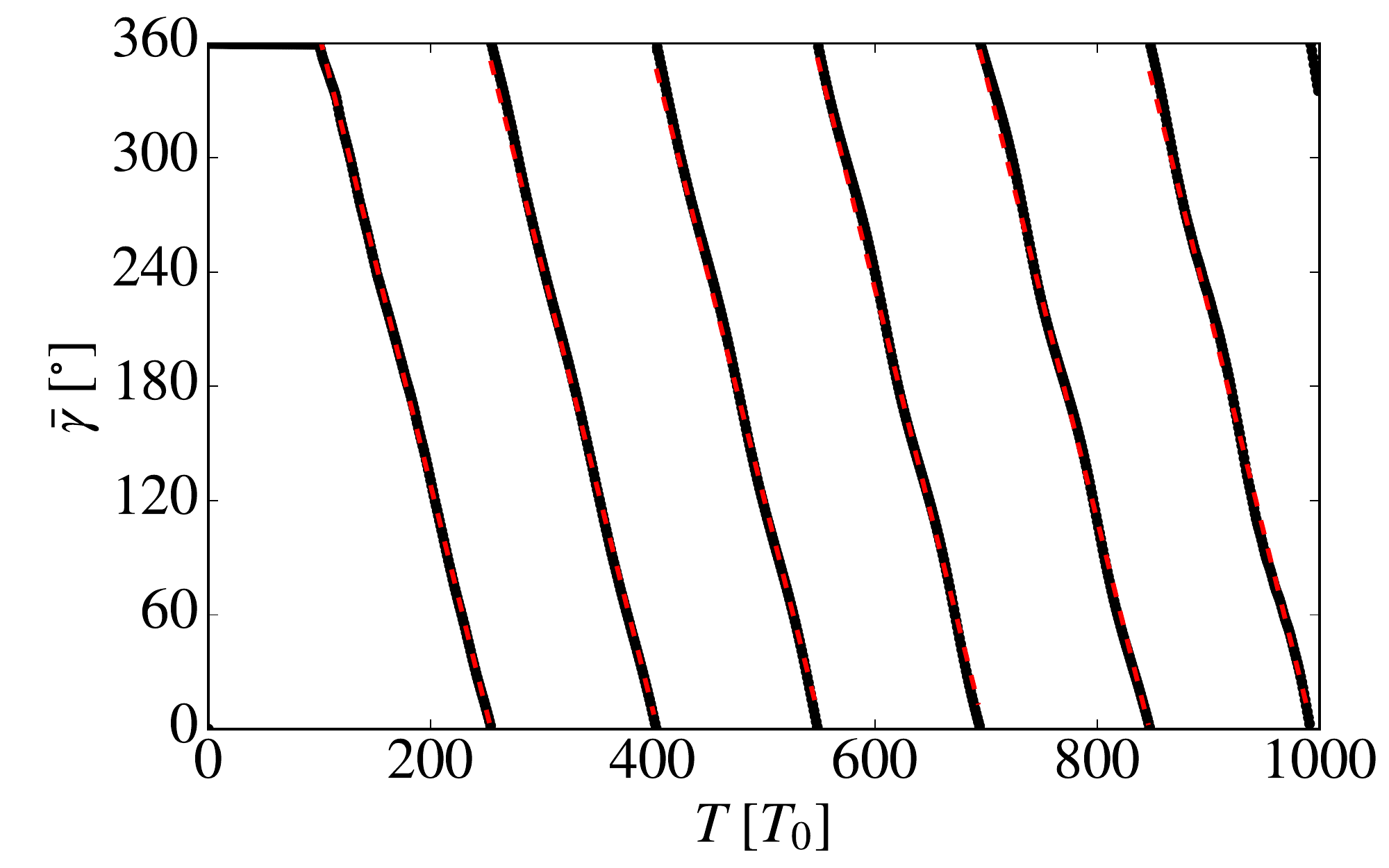}
        \caption{\textit{Top -} Angular momentum  $|\textbf{L}_z|$ (black), $|\textbf{L}_{\perp}|$ (blue), $|\textbf{L}_{x}|$ (red) and $|\textbf{L}_{y}|$ (green) for the Newtonian disc. \text{Bottom -} Mean precession angle $\gamma$ as a function of time for the Newtonian disc and the corresponding best linear fit showing $\omega_{\xi} = 0.04$.}
\label{fig:Delta_L}
\end{figure} 

In Fig. \ref{fig:angles} we plot the space-time diagram of misalignment angle $\beta$ and precession angle $\gamma$ for the Newtonian disc. The disc undergoes nearly perfect rigid body precession, as evidenced by the radially independent precession angle $\gamma$ while the inclination angle quickly decays from its initial value $\beta = 10^{\circ}$. 

Since the entire disc evolves approximately like a rigid body, for simplicity we perform a 1D analysis using the total angular momentum. In Fig \ref{fig:Delta_L} we plot each component of the total disc angular momentum $L_x$ (solid red line), $L_y$ (solid green line), $L_z$ (solid black line) as well as the perpendicular component  $L_{\perp} = \pm \sqrt{L_x^2 + L_y^2}$ (solid blue lines). We normalize $L_z$ by its value at $t = 100$ and all other values by $L_x$ at this same time. The $L_z$ angular momentum is conserved, save for a small fraction due to mass loss in the disc due to accretion. We perform a linear fit and find $L_{\perp} \sim e^{-t/\tau_{\perp}}$, with $\tau_{\perp} \approx 170$. By comparison, the time-scale for decay of $L_{z}$, driven by mass loss in the disc, is $\tau_z \approx 5 {} 300$. The $L_x$ and $L_y$ components undergo sinusoidal oscillations, out of phase by $\pi/2$, as expected for a precessing system. 

In the lower panel of Fig \ref{fig:Delta_L} we plot the mean precession angle (black points) and a linear fit, $\bar{\gamma} = \omega_{\xi} t $ which we fit using linear regression and find $\omega_{\xi} = 0.04$. Decomposing the angular momentum into its Fourier modes, we find a dominant mode with amplitude $\hat{L}_x = 53$ and frequency $\omega_{\xi} = 0.04$, driving the rigid body precession. By comparison, the angular momentum averaged precessional frequency (\ref{eq:fbar}) estimates $\bar{\omega} = 0.05$. In addition, there is a weaker, $\hat{L}_{\perp} = 5.4$ mode with frequency $\omega_{\eta} = 0.015$ mode. This corresponds to apsidal precession, expected from the coupling between the Lense-Thirring term and gravitational potential. The angular momentum averaged apsidal frequency $\bar{\eta} = 0.015$ when averaged over the radial range $4. \leq r \leq 16$. In fact, we can see from the space-time diagram for $\beta$ that the inner and outer disc are nutating out of phase by $\pi/2$ with the cutoff near $r \sim 20$. This suggests that apsidal precession in the inner disc drives waves in the outer disc, where its own apsidal precession is weaker. Though there is some apsidal motion (nutation) the amplitude is small compared to the precessional motion and we conclude the disc behaves to first order like a rigid body, weakly damped harmonic oscillator. The precessional frequency is the angular momentum weighted average of the precessional frequency of the individual rings and the damping scale is set by the gas viscosity. There is a very weak nutational motion, but it is subdominant to the precession.

\subsubsection{Gravitational Potential}
\label{sec:low_potentials}

\begin{figure}
                \centering
                \includegraphics[width=0.48\textwidth]{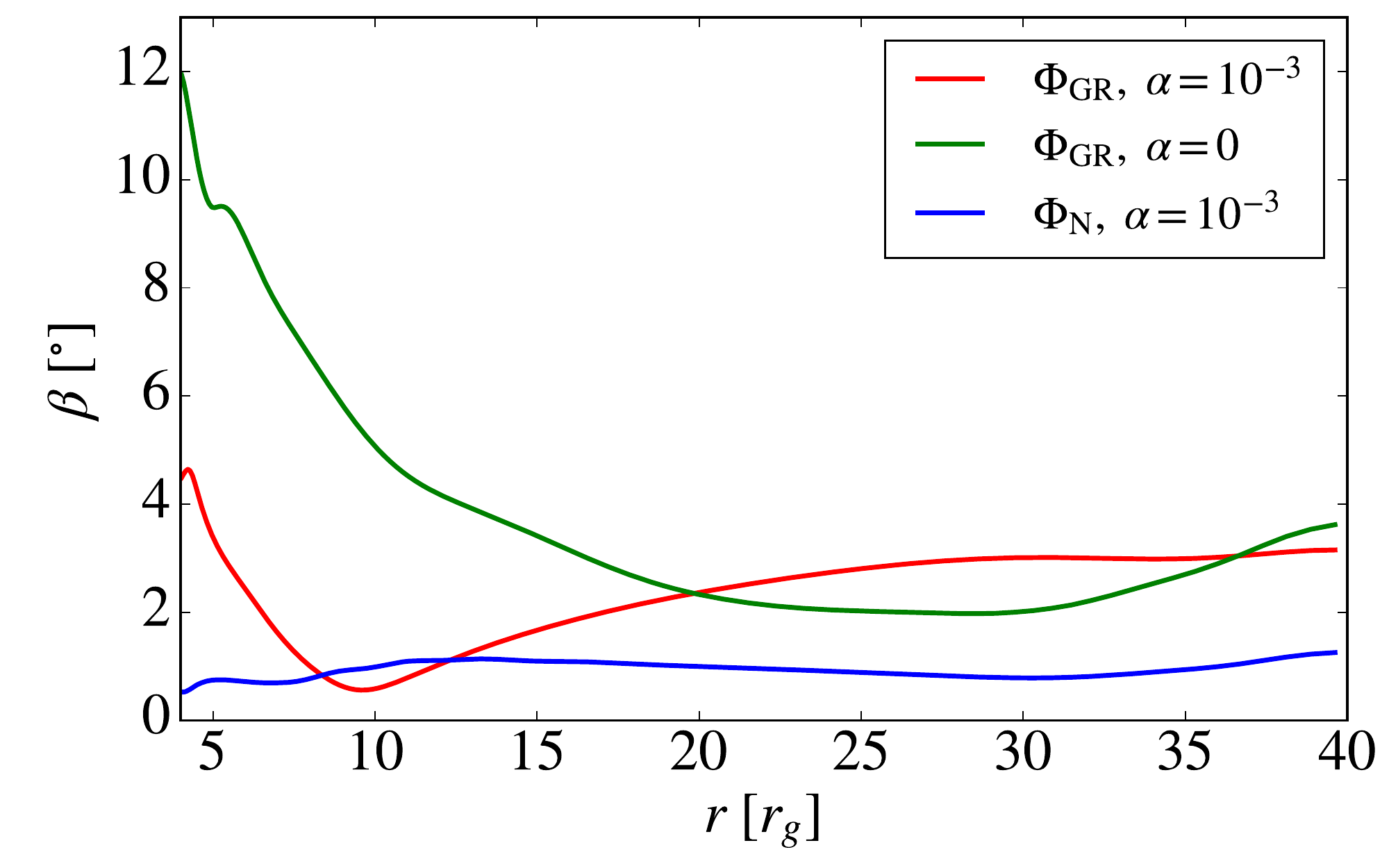}
        \caption{Axially averaged disc inclination $\beta$ for the Newtonian $\Phi_N$ (blue line), inviscid $\Phi_{GR}$ (green) and low viscosity (red) discs. The correct inner disc structure for the bending regime is reached in the low viscoisty disc.}
\label{fig:apsidal_term}
\end{figure} 

\begin{figure}
                \centering
                \includegraphics[width=0.48\textwidth]{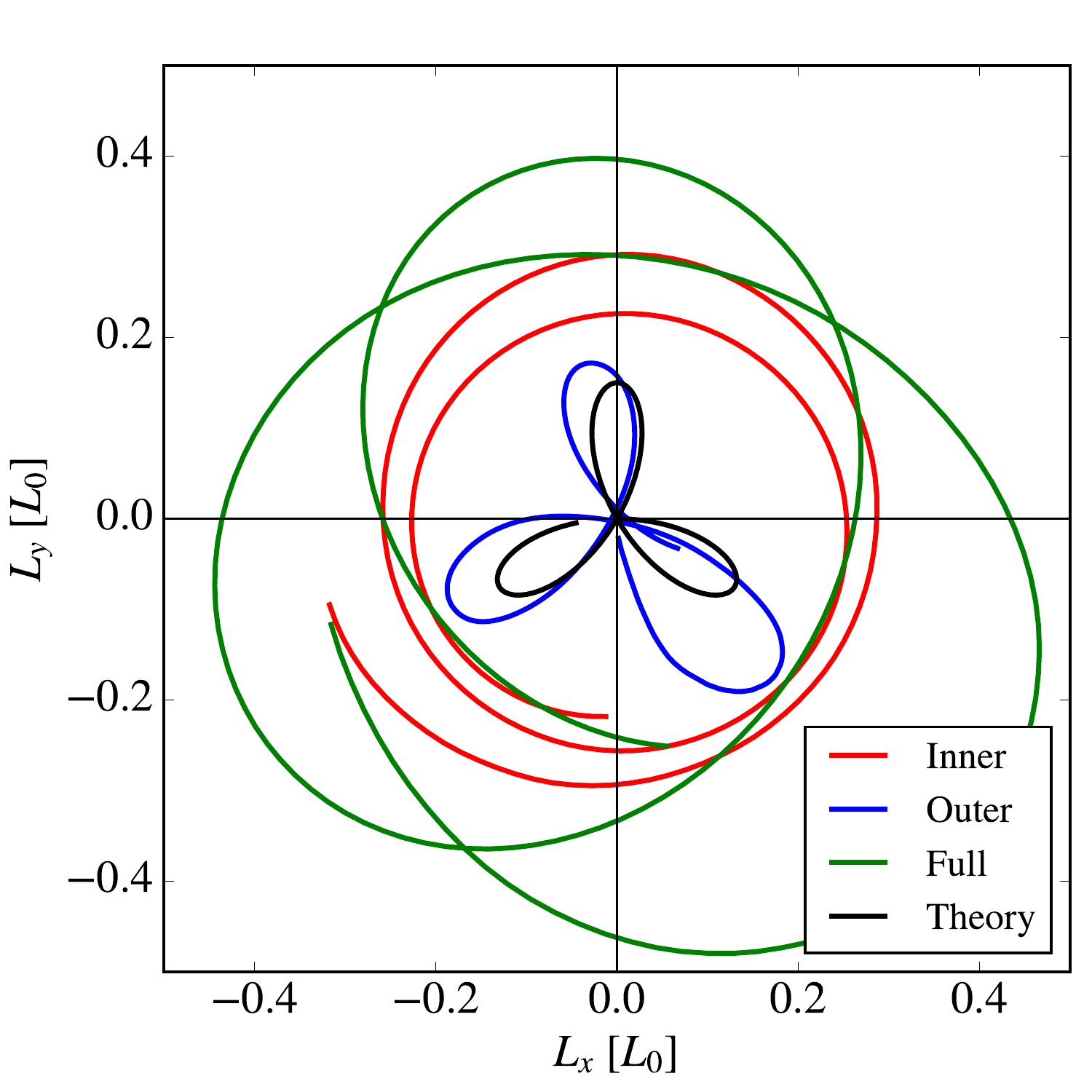}
        \caption{Evolution of angular momentum of the inner (red), outer (blue) and total disc (green) inviscid disc. We plot an analytical model for a precessing and nutating top with relative frequencies $\eta = 3 \xi$ (black), which approximately reproduces the motion of the outer disc.}
\label{fig:spirograph}
\end{figure} 

Before considering the dynamics of discs driven a pseudo-Newtonian potential, in Fig. \ref{fig:apsidal_term} we compare disc structures for the Newtonian model (blue line) and inviscid (green line) and low viscosity discs (red line) with pseudo-Newtonian potential (\ref{eq:GRPotential}) at $t = 500$. Firstly, as shown by Nealon et al. (2016), the effective GR potential is crucial to capture the qualitatively correct inner disc structure, namely the dip in the inclination at $r \approx 9 r_g$. Our low viscosity model is in good agreement with Nealon et al. (2016) (see their Figure 5). Our inviscid disc (green line) resembles the low viscosity disc at early times, before any angular momentum exchange has taken place. However, the inviscid disc does not develop the dip because angular momentum cannot be effectively exchanged. The Newtonian inner disc is aligned with the black hole spin, whereas including the GR correction induces the innermost part of the disc to remain misaligned. Qualitatively this behaviour was predicted in the linear regime for discs in the bending wave regime by Lubow, Ogilvie and Pringle (2002).

In terms of dynamics, GR correction to the gravitational potential adds a visibly stronger nutation to the damped harmonic motion of the disc. Intuitively, when $a = 0$ there is no apsidal motion in a Newtonian potential whereas in the GR potential this is not the case. In the inner parts of the disc, the nutation effect is small, whereas in the outer parts of the disc the amplitude is comparable to the disc inclination. In Fig. \ref{fig:spirograph} we plot the evolution of the perpendicular components of the angular momentum in the inner (red), outer (blue), and full disc (green). We define the outer disc, $20.7 \leq r \leq 40.0$ as the largest subdisc with nutation amplitude equal to the disc inclination i.e the part of the disc where nutation is dominant and the inner disc $r < 20.7$ The nutation causes the inclination angle $\beta = 0$ at some times, at which point the twist angle is undefined. We therefore work with the angular momentum vector rather than the Euler angles.  

A Fourier analysis shows that modes $\omega_{\xi} = 0.03, 0.06$ and $\omega_{\eta} = 0.09$ dominate the dynamics during $100 \leq t \leq 500$. In the inner disc only the $\omega_{\xi} = 0.06$ mode is excited and follows the motion of a damped, precessing top. The precession frequency corresponds to $\bar{\xi}$ over the range $4 \leq r \leq 30.$. In the outer disc, the modes $\omega_{\xi} = 0.03, 0.06$ have roughly equal amplitude, $\hat{L}_x = 30, 20$ respectively. In addition, the $\omega_{\eta} = 0.09$ has amplitude $\hat{L}_{\perp} = 20$. These modes correspond to $\bar{\eta}$ and $\bar{\xi}$ over the range $8.6 \leq r \leq 40.$ and $10.6 \leq r \leq 40.$ respectively. We interpret this to mean the dynamics is driven by waves from the inner and outer disc edge. Each propagates for a distance $\sim 30 r_g$ corresponding to the distance a wave of dimensionless speed $c_s/2v_{\phi}$ can propagate in a time $t \sim \bar{\eta}^{-1}$. 

The outer disc motion can be neatly described as a nutating, precessing top with
\begin{subequations}
\begin{equation}
L_x = L_{\perp} \cos(\bar{\eta} t) \cos(\bar{\xi} t)
\end{equation}
\begin{equation}
L_y = L_{\perp} \cos(\bar{\eta} t) \sin(\bar{\xi} t)
\end{equation}
\end{subequations}
In our case $\bar{\eta} \approx 3 \bar{\xi}$, resulting in a rhodenea curve with three petals (black line).

We see dissipation on time-scales $\tau_{\perp} \approx 470$, longer than for the Newtonian case. As precessional effects are \emph{weaker} in this case, as evidenced by the lower precession rate, we expect any dissipation to also occur on longer time-scales.            
  
\subsubsection{Viscous Effects}
\label{sec:low_viscosity}

\begin{figure}
                \centering
                \includegraphics[width=0.48\textwidth]{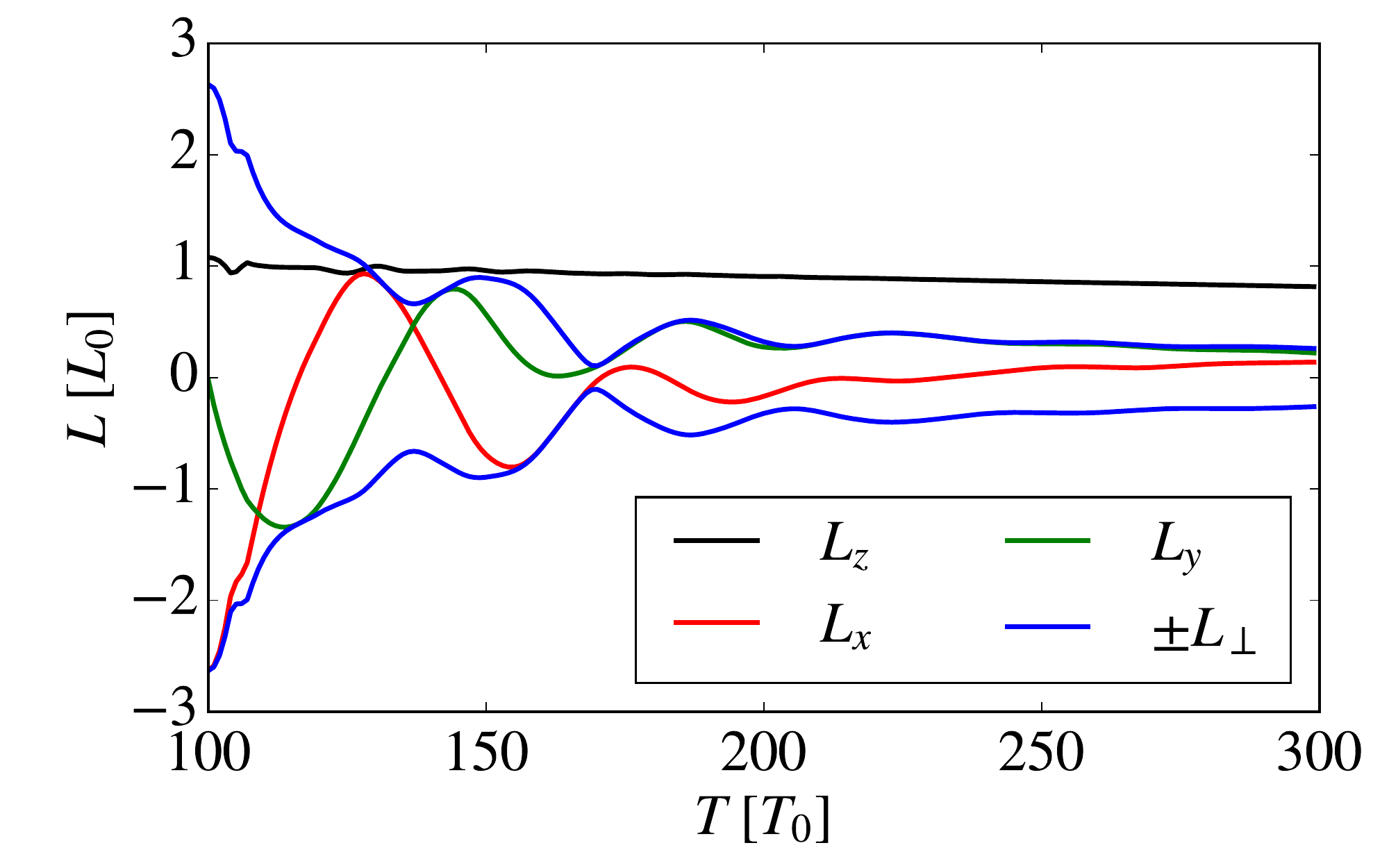}
        \caption{Angular momentum components in the range $4 \leq r \leq 10$ for the high viscosity disc showing the near critical damping as precession amplitude decays after $\sim 2$ oscillations.}
\label{fig:L_i10_alpha2}
\end{figure} 

We consider low viscosity, $\alpha = 10^{-3}$, and high viscosity, $\alpha = 10^{-2}$, discs, where viscosity is calculated explicitly using an $\alpha$ prescription. We have discussed the inviscid cases extensively in the previous sections. When viscosity is weak, the qualitative behaviour is unchanged with the disc evolving like a rigid body undergoing precession and nutation. The disc modes are identical to the inviscid case with $\omega_{\xi} =$ 1.8 and 5.4 and $\omega_{\eta} = 7.2$ in the interval $100 \leq t \leq 300$ and $\omega_{\xi} =$ 1.7 and 7.2 and $\omega_{\eta} =$ 9.0 in the interval $300 \leq t \leq 500$  The dissipative effects are enhanced, with a viscous time-scale $\tau_{\perp} \approx 300$, roughly 2.5 shorter than the inviscid case.

The dynamics of the high viscosity case is qualitatively different. In Fig. \ref{fig:L_i10_alpha2} we plot the total angular momentum in the range $4 \leq r \leq 10$. The disc undergoes (near-)critically damped oscillations, undergoing only two precessions before the $L_x$ angular momentum changes sign before decaying exponentially to zero. The disc misalignment decays $i \rightarrow 0$ on a timescale $\tau \approx 240$, shorter than the precession time $2 \pi \omega_{\eta}^{-1} \approx 100$. Critical damping corresponds to the case $\tau = 2 \pi \omega_{\eta}^{-1} $, in analogy with the classical harmonic oscillator. We therefore expect that for $\alpha \gtrsim 4 \times 10^{-2}$ any precession will be explicitly damped out, with our case approaching the critically damped case. The disc damps on a short enough time-scale that precession is not transmitted to the furthest parts, $r \gtrsim 20$ parts of the disc.

\subsection{Strongly Misaligned Discs}
\label{sec:high_inc}

\begin{figure*}
                \centering
                \includegraphics[width=\textwidth]{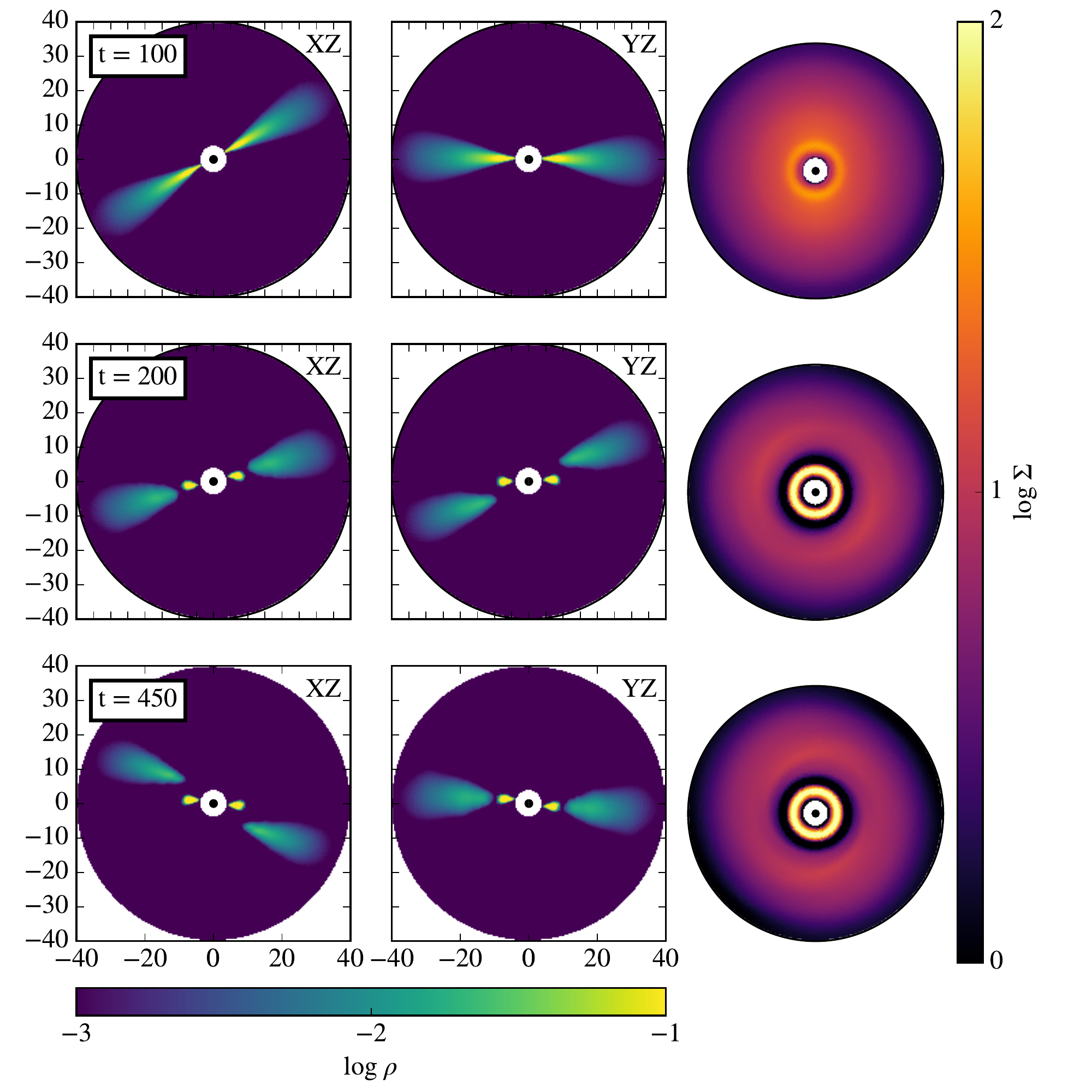}
        \caption{Density $\rho$ and surface density $\Sigma$ for $i=30^{\circ}$ inviscid disc}
\label{fig:summary_i30}
\end{figure*} 

When discs are strongly misaligned, $i \gtrsim 30^{\circ}$, they undergo a short transient phase before breaking at a radius $r_{\rm{break}}$. The inner, $r \lesssim r_{\rm{break}}$, and outer, $r \gtrsim r_{\rm{break}}$, sub-discs behave quasi-independently and evolve according to the same physics, albeit on different time-scales because of their relative distance to the black hole. We focus our investigation on the effects of disc thickness (Section \ref{sec:high_temp}), viscosity (Section \ref{sec:high_viscosity}), the shape of the gravitational potential (Section \ref{sec:high_potential}) and inclination angle (Section \ref{sec:high_inclination}).

In Fig. \ref{fig:summary_i30} we plot the density (left panels) in the XZ and YZ planes and surface density at three representative times in the motion of the inviscid disc. The $t = 100$ panel shows the disc the moment the Lense-Thirring torque is turned on. After 10's of inner disc orbits, the disc breaks at $r \approx 9 r_g$, where a gap in the surface density forms. The system then effectively behaves like two inviscid, low inclination discs with each sub-disc precessing at different frequencies. The inner disc precesses at a rate $\omega_{\rm{in}} = 0.28$ and the outer disc $\omega_{\rm{out}} = 0.015$. The final time at $t = 450$ shows the system after the outer disc has undergone approximately half a precession cycle.  

We may estimate the radius of the disc breaking as follows. The time-scale for the break to occur will be of order $t_{\rm{break}} \sim r_g/c_s$. If the disc is to break, on this time-scale, the inner disc should precess a full rotation thus $\bar{\xi} \approx 2 \pi/t_{\rm{break}}$, which defines the upper radius cutoff of the integral $r_2$ in equation (\ref{eq:GR_xi}). Waves from this outer cutoff travel both inwards and outwards at equal speeds, so the part of the disc that is causally connected and acting as a rigid body should extend from $r_{\rm{in}} \lesssim r \lesssim 2 r_2 - r_{\rm{in}}$, defining the breaking radius $r_{\rm{break}} \approx 2 r_2 - r_{\rm{in}}$. 

In this case, the sound speed $c_s = 0.05$ defines a breaking time $t_{\rm{break}} \approx 20$. This is consistent with when we see the disc breaking after we turn on the Lense-Thirring term. This timescale yields a mean precession rate $\bar{\xi} \approx 0.31$. By contrast, Fourier decomposition yields a dominant mode in the inner disc of $0.30$. Integrating over the disc, we find an upper cutoff $r_2 \approx 6.5$, yielding a breaking radius $r_{\rm{break}} \approx 9$. By comparison, from the surface density profile we find a gap in the annulus $9.1 \lesssim r_{\rm{gap}} \lesssim 11.1$, consistent with this estimate of $r_{\rm{break}}$. 

A better estimate for the breaking radius is derived from the inner disc precession frequency. Assuming $\bar{\xi} = 0.29$ is known, we can estimate $t_{\rm{break}} \approx 22$ and $r_2 \approx 6.6$. This yields an estimate of $r_{\rm{break}} \approx 8.6$.           

The rate of angular momentum loss $\tau_{\perp} \approx 3200$ is roughly ten times longer than for the low inclination disc. By breaking, each sub-disc can precess closer to its natural frequency and minimize internal damping. In particular, radial waves driven at the inner radius apsidal frequency are now driven from the inner part of the \emph{outer} sub-disc and therefore act on longer time-scales.  

\subsubsection{Disc Thickness}
\label{sec:high_temp}
We consider disc thicknesses in the range $0.025 \leq h/r \leq 0.1$. For the thicker discs, $h/r = 0.1$, the inner and outer parts of the accretion disc evolve independently. Disc metrics characterizing their dynamics, such as the inclination and precession angle vary discontinuously with radius. Though the surface density profile does decrease near such discontinuities, there is no explicit gap opening and diffuse gas still fills this region. Further our time-scale estimate based on the sound speed estimates a breaking radius at $r_{\rm{break}} \lesssim 5$, which likely cannot be resolved within our domain especially given that the surface density drops at the very inner edge. Because the break is less apparent in this case, we focus our analysis on the thin disc cases where we see a clean break.  

For thinner discs, $h/r = 0.025$, the disc evolution is qualitatively unchanged from the fiducial case. From the reduced sound speed, $c_s = 0.025$ we estimate $t_{\rm{break}} \approx 40$, the precession rate $\bar{\xi} \approx 0.16$, $r_2 \approx 9.2$ and a breaking radius $r_{\rm{break}} \approx 14.4$. From the surface density plot we see the gap actually opens at $11.1 \lesssim r_{\rm{gap}} \lesssim 13.2$, so our estimate accurate to better than $10\%$. The estimate is improved if we use the observed value of the mean precession rate $\omega_{\xi} = 0.17$ which yields $r_2 = 8.6$ and $r_{\rm{break}} \approx 13.2$, at the outer range of our disc gap. The precession rate of the inner disc thus seems to be a better indicator of the breaking radius than the sound speed.            

\subsubsection{Viscosity}
\label{sec:high_viscosity}
We consider discs with a low ($\alpha = 10^{-3}$) and a high ($\alpha = 10^{-2}$) viscosity. As in the low inclination case, the low viscosity disc is qualitatively the same as the inviscid case. The disc breaks and forms a gap between $9.9 \leq r \leq 11.1$, an increase of $\sim 10\%$ in the breaking radius. The $L_{\perp}$ angular momentum decays on a time-scale $\tau_{\perp} \approx 300$, roughly $30\%$ faster than the inviscid case. The inner disc has modes $\omega_{\xi} = 0.39$ and $\omega_{\eta} = 0.39$, corresponding to $\bar{\xi}$ over $4 \leq r \leq 6.2$ and $\bar{\eta}$ over $4 \leq r \leq 16.2$. Our estimate from the previous section would estimate $r_{\rm{break}} \approx 8.2$, an underestimate of the measured value of $\sim 20 \%$.   

The high viscosity case is qualitatively different. The disc breaks at $r_{\rm{break}} \approx 6.6$ and precesses for $\sim 20$ orbits  but does not form a clear gap. Matter accretes over this region, growing the inner sub-disc to $\sim 8.7$ before finally merging with the outer sub-disc. At late times the disc reaches a stationary, warped state with an inner part with $\beta \sim 0^{\circ}$ in the inner $r \lesssim 8$ and an outer disc with $\beta \sim 10^{\circ}$. These agree with our expectation of radial tilt oscillations at the inner disc edge. The inner disc mode $\omega_{\xi} = 0.63$ allows us to estimate a breaking radius $r_{\rm{break}} \approx 6.2$, slightly below the observed breaking radius. As with our other cases, we tend to \emph{under-estimate} breaking radii for more viscous cases. This is expected as viscous forces allow the disc to communicate torques across the disc and better evolve like a rigid body.        

\subsubsection{Gravitational Potential}
\label{sec:high_potential}

\begin{figure}
                \centering
                \includegraphics[width=0.45\textwidth]{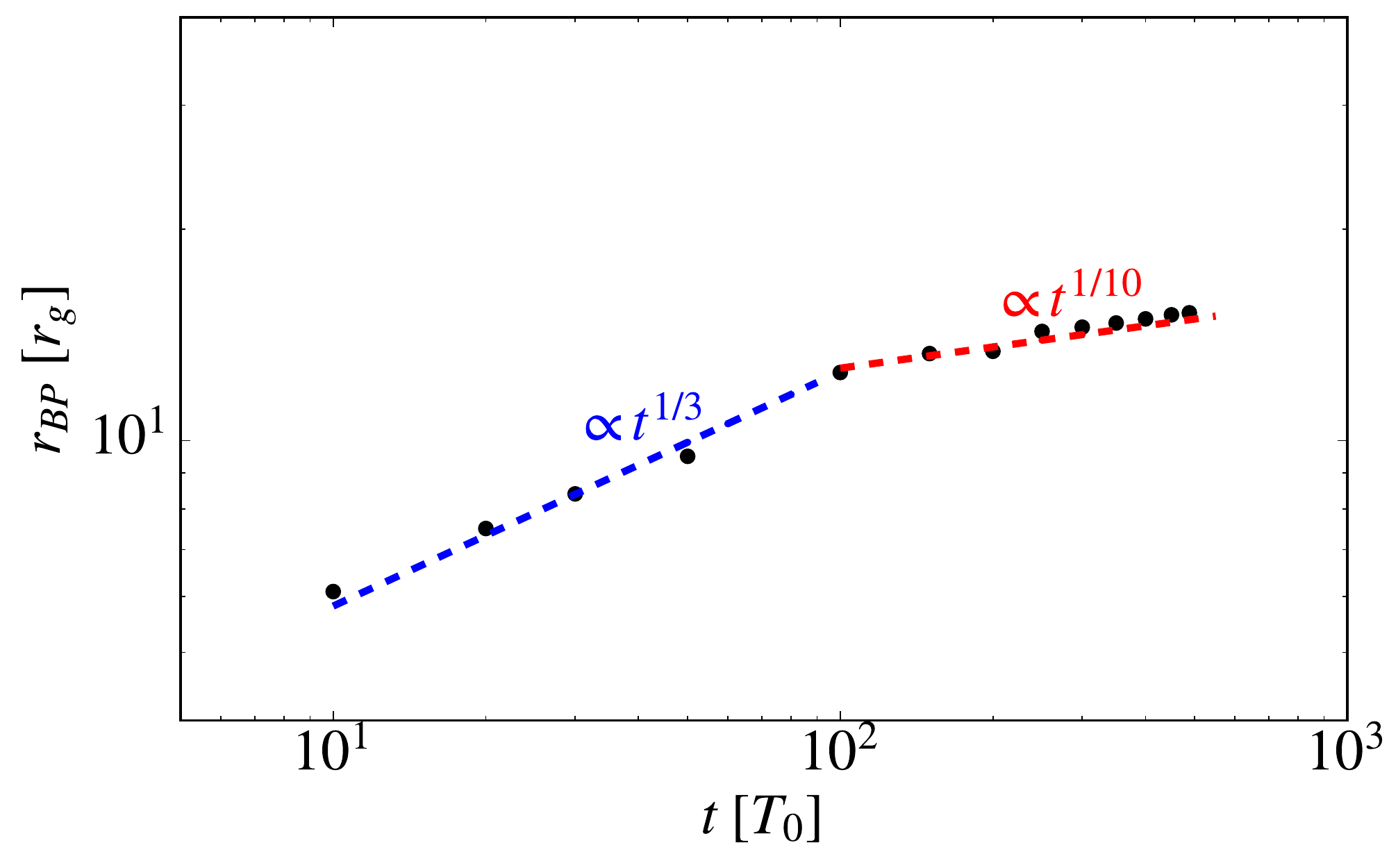}
        \caption{Bardeen-Petterson alignment radius as a function of time. At early times it scales like $\propto t^{1/3}$, in agreement with the precession time-scale $\xi^{-1} \sim r^3$. At late times the alignment radius grows like $\propto t^{1/10}$.}
\label{fig:BP-rad}
\end{figure} 

We have chosen a case with Einstein potential as our fiducial case, because as we have argued in Section \ref{sec:low_inc}, it is required to produce the correct inner disc structure, including tilt oscillations, to generate the correct disc dynamics. To further demonstrate this, we use our fiducial high-inclination disc parameters but now in a Newtonian potential (\ref{eq:NewtonPotential}).

Unlike all other strongly misaligned models, the inner disc aligns with the midplane as prescribed by the Bardeen-Petterson picture. In Fig. \ref{fig:BP-rad} we plot the time evolution of the Bardeen-Petterson radius $r_{\rm{BP}}$ and show its empirical scalings. The innermost disc, $r \lesssim 12$ aligns after $\sim 100$ inner disc orbits during which the Bardeen-Petterson alignment radius $r_{\rm{BP}}$ grows at the rate $r \sim t^{1/3}$, in agreement with the precession time-scale $\xi^{-1} \sim r^{3}$. As inner disc aligns, accretion from the outer disc slows and the alignment proceeds along a much flatter $r \sim t^{1/10}$. During the alignment phase it undergoes precessional and nutational motion $\omega_{\eta} = \omega_{\xi} = 0.12$. This precessional frequency is close to the mean $\bar{\xi} = 0.11$ for early times $100 \leq t \leq 200$.  Meanwhile the outer disc undergoes Lense-Thirring precession, $\omega_{\xi} = 0.005$. The precessional motion prevents most accretion from the outer to the inner disc, save for accretion streams where they are both in the black hole spin plane.  

This case clearly demonstrates the need for an Einstein potential to properly model the disc dynamics. In a purely Newtonian potential tilt oscillations are absent and the inner disc aligns aligns due to the Bardeen-Petterson effect. This occurs despite the disc being formally inviscid due to radial pressure gradients. By contrast, in the case of a low inclination disc the pressure gradient forces are not so great and alignment does not occur on the local Keplerian time-scale. The disc behaves more like a rigid body and aligns on a disc-averaged viscous time scale.  

\subsubsection{Inclination Effects}
\label{sec:high_inclination}

\begin{figure*}
                \centering
                \includegraphics[width=\textwidth]{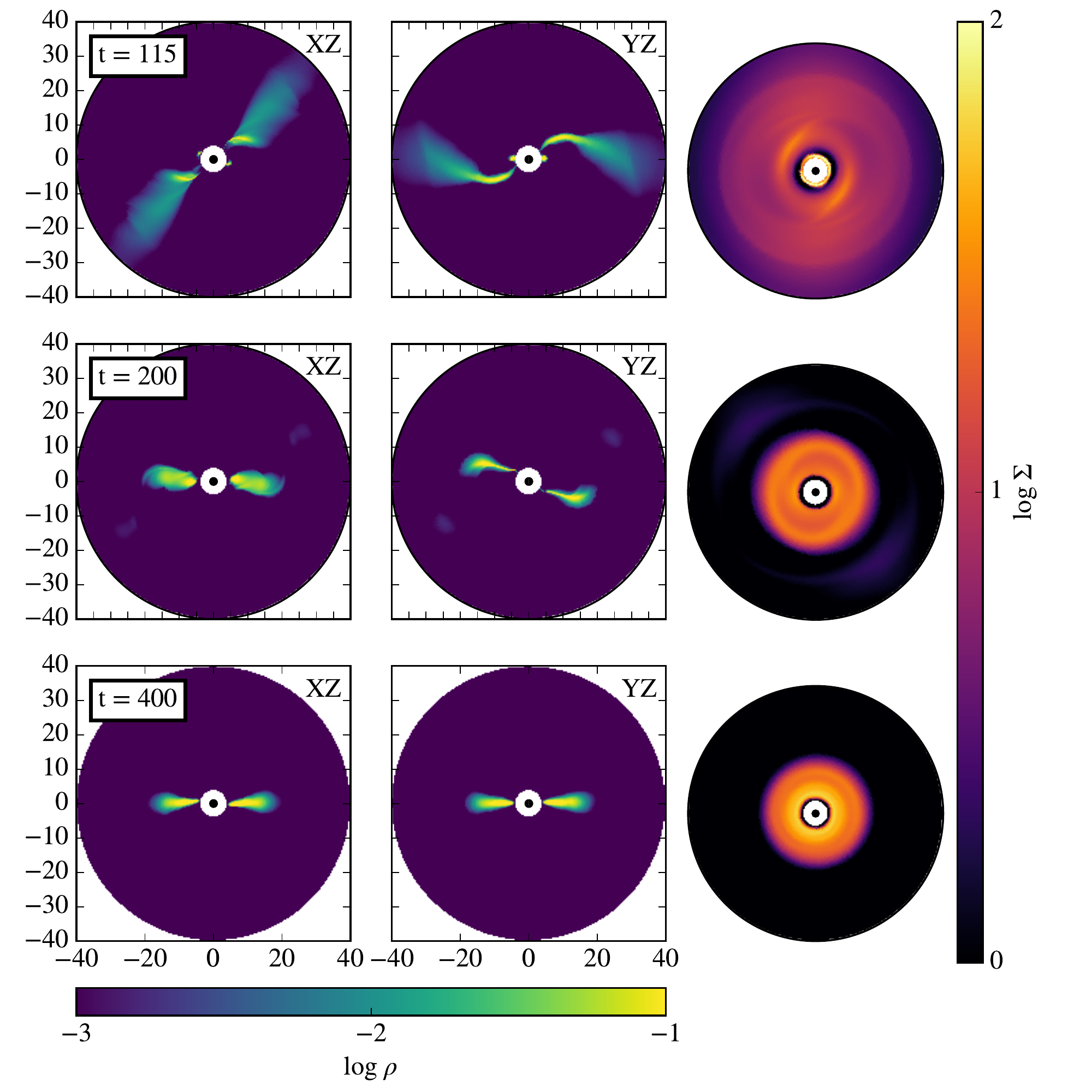}
        \caption{Density $\rho$ and surface density $\Sigma$ for $i=45^{\circ}$ high misalignment disc}
\label{fig:summary_i45}
\end{figure*} 

We consider our fiducial inviscid disc with an inclination of $i = 45^{\circ}$, illustrated at representative times in Fig. \ref{fig:summary_i45}. The greater inclination angle enhances the effects of the Einstein potential and leads to more violent dynamical evolution during the initial transient. Prior to turning on the Lense-Thirring term the disc reaches a steady state with a visible warp in the innermost $\sim 5$. This is a region with high pressure gradients and after turning on the Lense-Thirring term the disc breaks nearly instantly at $r \sim 5.4$. This innermost sub-disc precesses in the clockwise direction, as expected from the low inclination models, and at a high frequency of $\omega = 0.94$ (top panel, $t = 115$). Meanwhile, the outer disc, initially at a higher inclination angle also begins to precess, but at a lower frequency. As this outer disc precesses, its inclination begins to decreases and it eventually matches the inclination of the inner disc, thereby allowing significant mass transfer to the inner disc as misaligned angular momentum between the inner and outer disc annihilate and allow gas to accrete. Eventually, the inner disc with high precession rate fully accretes onto the black hole around $t = 150$, leaving behind a lower inclination $i \lesssim 10^{\circ}$, accretion disc that evolves much like our lower inclination runs. 

Meanwhile similar dynamics is at play, although less violent, at $r \sim 20$, where an additional disc breaking occurs (center panel, $t = 200$). The break is characterized by a discontinuous jump of $\Delta \beta = 10^{\circ}$ in the inclination angle. At such large radii, the outermost part of the disc does not have time to precess more than a quarter period before gas is depleted and the system is composed of only middle part of the disc (bottom panel, $t = 400$). This is however consistent with high inclination simulations by Nealon et al (2016) and Lisaka et al (2019) which found multiple disc breaks in their highest inclination, highest resolution simulations.    

This case illustrates that at high inclination angles the system can undergo interesting transient behaviour i.e multiple disc breaks, large disc warps, high accretion rate, etc... At late times the systems tend to settle to one of the more stable, lower inclination scenarios. The final state, single low inclination disc vs. broken inner/outer disc seems to depend on whether the accretion rate during the transient phase leaves sufficient matter in the disc.

\section{Discussion}
\label{sec:discussion}

We are the first to report disc breaking in grid based HD simulations of warped discs. In order for discs to break we find three necessary conditions 1) A pseudo-Newtonian potential of the form (\ref{eq:GRPotential}) 2) strong misalignment $i \gtrsim 30^{\circ}$ 3) sufficiently thin $h/r \lesssim 0.05$. We found that when a Newtonian potential is used, discs Bardeen-Peterson align and do not break. Likewise, when the misalignment is weak, discs rigid body precess and do not break. If the scale-height is too large the inner and outer discs smoothly transition inclinations and do not evolve independently as is characteristic of a break. 

These results are in agreement with the growing suite of misaligned disc simulations. The MHD simulations of Krolik et al (2015) found no disc breaking as they used a Newonian potential and their discs were only weakly misaligned $i \approx 10^{\circ}$. The SPH simulations of Nealon et al. (2015) with pseudo-Einstein potential found disc breaking for $i \gtrsim 30^{\circ}$ but none for $i \lesssim 15^{\circ}$. GRMHD suggests that criteria (2) and (3) are also crucial in resolving the tearing question. White et al. (2019) found no tearing but studied discs with $i \leq 24^{\circ}$. Liska et al (2019, 2020) found that weakly misalinned, thin discs ($i = 10^{\circ}, h/r = 0.03$) did not break while strongly misaligned $i \geq 45^{\circ}$ discs do break.

Despite agreement on the criteria for disc breaking amongst the simulations other features require further clarification. SPH simulations have found multiple disc breaks, with very narrow subdiscs $\Delta r \sim h$ whereas GRMHD has found $\Delta r \gg h$. Liska et al. (2020) speculated that radial tension in the field lines may help keep subdiscs together and account for the narrow subdiscs in (hydrodynamic) SPH. However we also find wide subdiscs despite the purely HD nature of the simulations. In grid based HD the numerical viscosity is fixed by the grid size, unlike in SPH where it will vary with the particle density. This suggests that as gaps form the numerical viscosity in these regions can become large favoing further disc breaking. 

We found we can approximate the disc breaking radius by comparing the angular momentum weighted precessional frequency and the sound crossing time scale. The crucial pieces of physics for breaking the disc are the Lense-Thirring torque and pseudo-Newtonian potential. The latter is chosen to correctly reproduce the apsidal precession frequency at large radii. We have shown that disc breaking is primarilly driven by the inner disc modes and therefore any quantitative predictions of breaking radii for non-GRMHD simulations based on these frequencies are suspect. A careful code comparison would be needed to determine how closely disc breaking and precession frequencies agree amongst the different methods. Our high inclination discs break at a similar radius to the $a = 0.9$ spin cases in Nealon (2015). However, when we apply our breaking criteria we systematically over-estimate the breaking radius they found for lower spin cases. They however used a different surface density and sound speed profile so comparison is difficult.

GRMHD simulations unambiguously find the innermost parts of efficiently cooled discs (small h/r) align with the black hole spin. Liska (2019, 2020) showed $h/r = 0.03$ and $i = 10^{\circ}$ discs align in the very inner $r \lesssim 5 r_g$ (2019) and increasing the misalignment to $i = 65^{\circ}$ for $h/r = 0.03$ breaks the alignment for some times while increasing the thickness to $h/r = 0.05$ for $i = 45^{\circ}$ prevents alignment (2020). By comparisson White et al. (2019) found no alignment for $0.1 \leq h/r \leq 0.2$. The picture for non-GR codes is less clear. Lower resolution SPH simulations (Nelson \& Papaloizou 2000) have found inner disc alignment. More recent simulations by Nealon et al. (2015) found agreement with their results but at higher resolution they found that the discs would instead break. Our grid based HD simulations begin to show alignment in only one case, the strongly misaligned Newtonian disc. In both our grid HD and the SPH work the inner radius $r_{\rm{in}} = 4 r_g$, well outside the ISCO for the high spin (a = 0.9) cases we have studied. Thus despite simulating a thin enough disc, $h/r = 0.025$, as the GRMHD work found discs breaking at $r_{\rm{BP}} \approx 5 r_g$ it is perhaps unsurprising that we have not found Bardeen-Petterson alignemnt in our work.

Observationally disc breaks may be important as precessing inner discs have been proposed as a mechanism for generating type-C low frequency QPOs (Ingram et al. 2009). To match observed frequencies, a precessing hot inner flow must be truncated at some radii by the inner edge of a cool disc. In our models, the disc breaks where the integrated precession time equals the sound crossing time of the sub-disc. By measuring QPO frequency, we can determine the precessional frequency of the inner sub-disc. We can then apply our criteria for disc breaking and derive an estimate for the breaking radius as a function of sound speed. These can be compared to reflection spectra modeling, which provide an estimate of the temperature. New reflection models are also beginning to account for the disc break, providing an indepedent check on the breaking radius (Abarr \& Krawczynski 2020). 

The HD models of this study are limited in several ways. We find that discs with a uniformly high viscosity ($\alpha = 10^{-2}$) damp on the precessional time scale. In moving on to MHD models, an important question is whether the effective viscosity can reach comparable levels locally but the disc can still precess globally. Furthermore, can viscosity become large in the inner disc, thereby allowing Bardeen-Petterson alignment in the inner disc, while allowing weakly damped precession in the outer disc? Likewise, an important issue is whether discs will break at radii comparable to HD models with comparable effective viscosities or even break at all. Apart from generating the MRI, magnetic fields also act to launch winds and jets. We have been careful to study cold discs that do not drive a wind. However significant angular momentum can be lost to jets and there may further complicate the picture as Liska et al (2018) have shown.

\section{Conclusions \& Future Work}
\label{sec:conclusion}
We have performed a series of hydrodynamics simulations including the effects of GR with both a gravitational potential and Lense-Thirring correction. For small misalignment angles we find discs exhibit the full range of behaviour seen in classical harmonic motion - precession, nutation and damping, driven by the Lense-Thirring, gravitational potential and viscosity respectively on the appropriate time-scales. For large misalignments, discs are found to tear with inner and outer sub-discs behaving quasi-independently, save for a small amount of mass and angular momentum transfer between them. We can estimate the breaking radius knowing the sound speed and angular momentum distribution, or alternatively the inner disc precession time-scale.

Disc viscosity plays a key role in the system evolution. When it is small, $\alpha \lesssim 10^{-3}$ inner disc precession is only weakly damped and exhibit long-lived precession. Increasing the viscosity $\alpha \gtrsim 10^{-2}$ makes inner disc precession short lived and the disc forms stationary warps. Both scenarios may be observable using iron line reflection models or QPOs and may constrain disc radius and precessional speeds. Given the importance of viscosity in determining the dynamics, future work will generate viscosity self-consistently via the MRI. We can then compute effective $\alpha-$viscosities and compare these models with hydrodynamic models. 

We, along with other authors have considered discs with high $i \gtrsim 45^{\circ}$ misalignments. A further unexplored question is the suitability of our initial conditions and whether such highly misaligned discs can be formed in-situ. Such studies may suggest an upper range above which discs can be considered highly transient and not expected to be observed. We already see evidence of this with the highest inclination discs losing angular momentum, accreting matter and reaching quasi-stationary states resembling our initial conditions for lower inclination discs.

\section*{Acknowledgements}
SD acknowledges useful discussions with Jim Pringle, Steve Lubow, J. J. Zanazzi and Alessia Franchini.  SD and CSR acknowledge the UK Science and Technology Facilities Council(STFC) for support under the New Applicant grant ST/R000867/1 and the European Research Council (ERC) for support under the European Union's Horizon 2020 research and innovation programme (grant 834203). 

This work was performed using resources provided by the Cambridge Service for Data Driven Discovery (CSD3) operated by the University of Cambridge Research Computing Service (www.csd3.cam.ac.uk), provided by Dell EMC and Intel using Tier-2 funding from the Engineering and Physical Sciences Research Council (capital grant EP/P020259/1), and DiRAC funding from the Science and Technology Facilities Council (www.dirac.ac.uk).


\appendix
Below we describe the details of our simulation setup. The main variable of interest in our investigation is the disc tilt $i$, which must be non-zero for Lense-Thirring precession to be active. 

We will refer to two coordinate systems - the lab frame will use unprimed coordinates, with Cartesian $(x,y,z)$ and spherical $(r,\theta,\phi)$. This corresponds to the coordinates of our simulation, and the system in which the black hole spin $\mathbf{J} = J_0 \hat{z}$. Relative to this coordinate system, we have a Keplerian accretion disc, inclined by an angle $i$ relative to $\mathbf{J}$. The disc frame will use primed coordinates, Cartesian $(x',y',z')$ and corresponding spherical coordinates $(r',\theta',\phi')$. 

\section{Initial Conditions}
\label{sec:app_ic}
\begin{figure}
                \centering
                \includegraphics[width=0.48\textwidth]{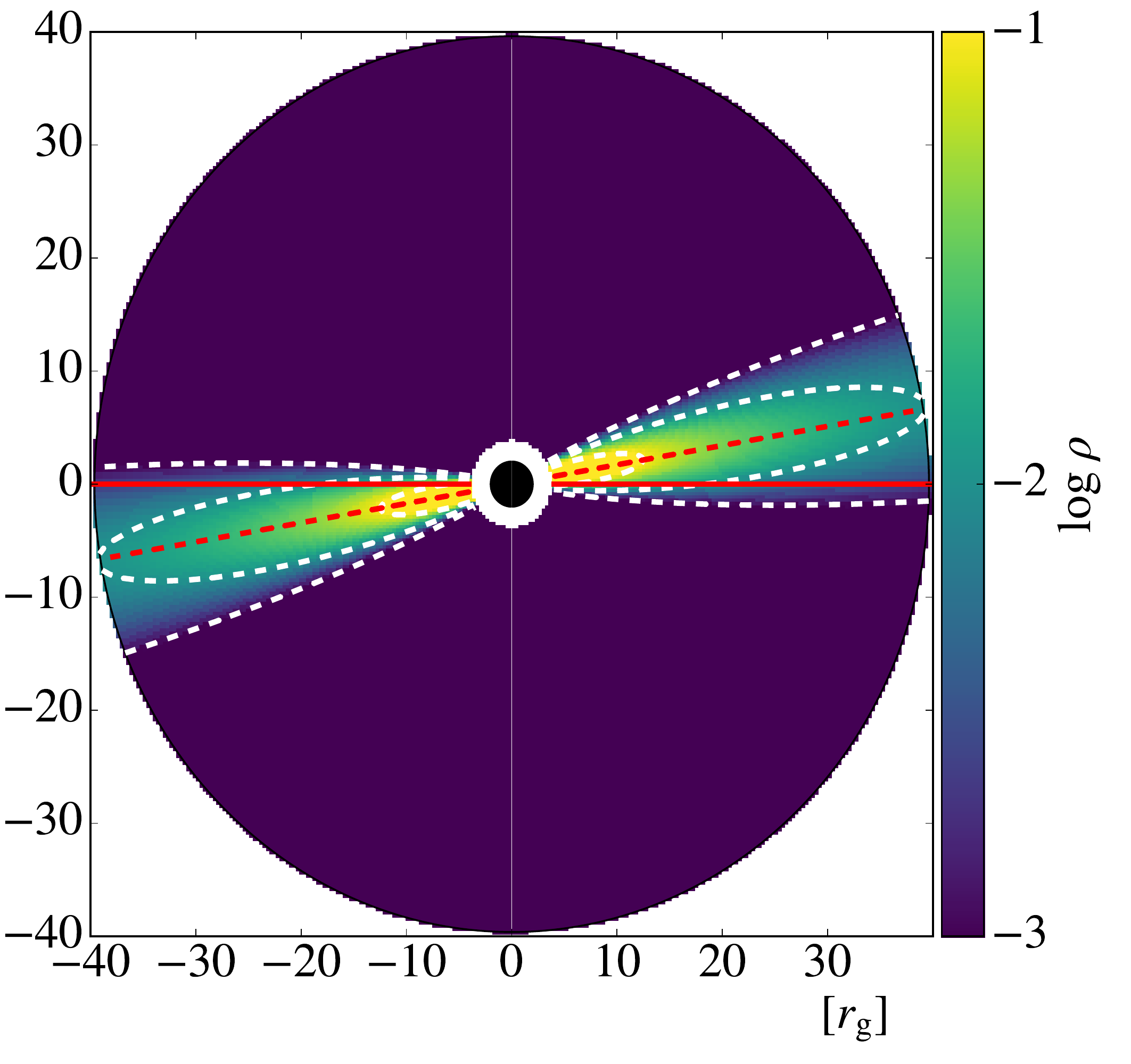}
        \caption{Density contours (black dashed lines and color contours) in the x-z plane. Dashed lines indicate factor of 10 changes in density. The red dashed line indicates the initial disc midplane and the red solid line the equatorial plane in the black hole frame.}
\label{fig:IC}
\end{figure} 

In the disc centered coordinate system we choose a vertically hydrostatic density profile
\begin{equation}
\rho = \rho_0(r'\sin \theta') \exp \left\{ -\frac{GM}{2c_s^2 r' \tan^2\theta'} \right\},
\label{eq:density}
\end{equation}
where $\rho_0 = \rho_*(r'\sin \theta')^{\alpha_{\rho}}$ is a powerlaw along the midplane that we set to $\alpha_{\rho} = 2$ and $\rho_* = 1.$ for simplicity. For velocity we set a purely Keplerian velocity profile along vertical cylinders
\begin{equation}
\mathbf{v} = \sqrt{\frac{GM}{r'\sin \theta'}} \ \hat{\phi'}.
\label{eq:velocity}
\end{equation}
We convert from disc coordinates to black hole coordinates by rotating by an angle $-i$ about the $\hat{y}$ axis. Explicitly we have the following coordinate transformation
\begin{subequations}
\begin{equation}
x' = x \cos i + z \sin i,
\end{equation}
\begin{equation}
y' = y,
\end{equation}
\begin{equation}
z' = -x \sin i + z \cos i.
\end{equation}
\label{eq:Cartesian_transform}
\end{subequations}
Expressing the density profile (\ref{eq:density}) in terms of Cartesian coordinates
\begin{equation}
\rho = \rho_* \exp \left\{ -\frac{GM}{2c_s^2} \frac{z'^2}{r\left( r^2 - z'^2\right)}\right\},
\end{equation}
where we have used $r' = r$ (i.e r invariant under rotations). We may then express the density profile in the unprimed coodinates. Likewise we convert the velocity profile (\ref{eq:velocity}) to unprimed coordinates. First convert to primed Cartesian coordinates
\begin{equation}
\mathbf{v} = -\sin \phi' \sqrt{\frac{GM}{r' \sin \theta'}} \ \hat{x'} + \cos \phi' \sqrt{\frac{GM}{r' \sin \theta'}} \ \hat{y'}.  
\label{eq:velocity_cp}
\end{equation} 
We then find
\begin{align}
\mathbf{v} = \sqrt{\frac{GM r^2}{\left( x'^2 + y^2\right)^{3/2}}} \ &\Bigg[ \sin i \sin \phi \ \hat{\theta} \nonumber  \\ &+  \Big( \cos i \sin \theta + \sin i \cos \theta \cos \phi \Big) \ \hat{\phi} \Bigg].
\end{align}

\section{Lense-Thirring Term}
\label{sec:LenseThirring}
In our setup $\mathbf{J} = J_0 \hat{z} = a (GM)^2/c^3 \ \hat{z}$ with $a$ the dimensionless spin parameter implemented in the code. Carrying out the vector products in (\ref{eq:gravitomagnetic}) we find
\begin{align}
\mathbf{h} = \frac{2J_0}{r^3} &\Bigg[ -3 \sin \theta \cos \theta \cos \phi \ \hat{x} \nonumber \\ &-3 \sin \theta \cos \theta \sin \phi \ \hat{y} + \left( 1 - 3 \cos^2 \theta \right) \ \hat{z} \Bigg].
\end{align}
Converting to spherical coordinates
\begin{equation}
\mathbf{h} = -\frac{J_0}{r^3} \Bigg[ 2 \cos \theta \ \hat{r} + \sin \theta \ \hat{\theta} \Bigg].
\end{equation}
We precalculate $\mathbf{h}$ and store the result in an array so the change in momentum $d\mathbf{p} = dt \ \rho \left( \mathbf{v} \times \mathbf{h}\right)$ can be computed at each half time-step as a standard user defined source term in \textsc{Athena++}.

\label{lastpage}

\end{document}